\documentclass[11pt,a4paper]{JHEP3}
\usepackage{amssymb}
\usepackage{euscript}
\def\l{\lambda}
\def\a{{\alpha}}
\def\b{{\beta}}

\def\l{\lambda}
\def\x{{\rm x}}
\def\k{\kappa}
\def\e{{\epsilon}}
\def\bra#1{\langle #1 |}
\def\ket#1{|#1 \rangle}
\def\0{\nonumber}
\def\sin{{\rm sin}}
\def\cos{{\rm cos}}
\def\det{{\rm det}}
\def\Det{{\rm Det}}
\def\log{{\rm log}}
\def\tr{{\rm tr}}
\def\Tr{{\rm Tr}}
\def\exp{{\rm exp}}

\newcommand\T{{\cal{T}}}
\newcommand\Q{{\cal{Q}}}
\newcommand\K{{\cal{K}}}

\newcommand\V{{\cal{V}}}

\newcommand\M{{\cal{M}}}
\newcommand\N{{\cal{N}}}
\newcommand\CP{{\cal{P}}}

\newcommand\ee{\end{eqnarray}}	 	
\newcommand\be{\begin{eqnarray}}
\newcommand\ba{\begin{array}}			
\newcommand\ea{\end{array}}
\newcommand\eeq{\end{equation}}	 	
\newcommand\beq{\begin{equation}}

\preprint{SISSA/100/03/EP\\\tt hep-th/0311198}

\title{Dressed Sliver solutions in Vacuum String Field Theory}
\author{L. Bonora, C. Maccaferri,\\
International School for Advanced Studies (SISSA/ISAS)\\
Via Beirut 2--4, 34014 Trieste, Italy, and INFN, Sezione di
Trieste\\
E-mail:   \email{bonora@sissa.it}, \email{maccafer@sissa.it} }

\author{P. Prester\\
Department of Theoretical Physics, Faculty of Science,
University of Zagreb\\
Bijeni\v{c}ka c.\ 32, p.p.\ 331, HR-10002 Zagreb, Croatia\\
E-mail: \email{pprester@phy.hr}}

\abstract{We consider a new class of solutions (dressed slivers) in
Vacuum String Field Theory, which represent D25--branes. For each 
dressed sliver we introduce
a deformation parameter and define a family of states
which are characterized by new abelian $*$--subalgebras.
We show that this deformation parameter can be used as a regulator:
it allows us to define for each such solution a finite norm and energy
density. Finally we show how to generalize these results to parallel
coincident and to lower dimensional branes.}
\begin{document}

\section{Introduction}

VSFT, \cite{Ras}, is a version of Witten's open SFT, \cite{W1}, that is
supposed to describe the  theory at the minimum of the tachyonic potential.
There is evidence that at this point the negative tachyonic potential exactly
compensates  for the D25--brane tension. No open string mode is expected to be
excited, so that the BRST cohomology must be trivial. This state can only 
correspond to the closed string vacuum. A time-dependent solution 
which describes the evolution from the maximum of the tachyon potential 
to such a minimum (a rolling tachyon), if it exists, would describe 
the decay of the D25--brane into closed string states. That such a 
solution exists has been argued in many ways; in particular this has
led to the formulation of a new kind of duality between open and closed
strings \cite{Senroll}.
Although there have been some attempts to describe such phenomena in a 
SFT framework \cite{Kluson}, no analytical control has been achieved so far.
\\
In this regard VSFT could play an important role. VSFT is a simplified 
version of SFT. The BRST operator ${\cal Q}$ takes a very simple form
in terms of ghost oscillators alone. It is clearly simpler to work in
such a framework than in the original SFT. In fact many classical 
solutions have been shown to exist, which are candidates for representing
D--branes (the sliver, the butterfly, etc.), and other classical
solutions have been found (lump solutions) which may represent
lower dimensional D--branes \cite{KP, RSZ1, RSZ2, RSZ3, okuda, GRSZ2, MT}. In
some cases the spectrum around such  solutions has been
analyzed,\cite{HKw, HM1, HM2, HM3, RSZ4, HK, David, Oka} and there are
hints that it provides the modes of a D--brane spectrum. However the responses
of VSFT are still far from being satisfactory. There are a series of nontrivial
problems left behind. Let us consider for definiteness the sliver solution.
To start with it has vanishing action for the matter part and
infinite action for the ghost part, but it is impossible to make a 
finite number out of them, \cite{Oku2}. Second, it is not at all clear, at least
in the operator formalism, whether
the solutions to the linearized equations of motion around the sliver
can accommodate all the open string modes (as one would expect if the sliver
has to represent a D25--brane). Third, the other Virasoro constraints on
such modes are nowhere to be seen.

In the face of this evidence, one can conclude either that VSFT is an
oversimplified theory which has lost too many degrees of freedom to 
describe string physics correctly, or that the sliver is not a very convenient 
solution to represent a D25--brane. In this paper we take the second 
point of view: we wish to find a solution to the
VSFT equation of motion which is more appropriate than the sliver
to represent the D25--brane of open string field theory.
On the other hand the sliver has many interesting properties: 
it is simple and algebraically appealing (it is a squeezed state), 
its Neumann matrix $CS$ commutes with the twisted matrices of the 
three strings vertex coefficients and the calculations involving
the sliver are relatively simple. Therefore in order to define a new 
solution we choose to stay as close as possible to the sliver. In practice
we start from the sliver and `perturb it' by adding to $CS$ a suitable 
rank one 
projector $P$. We can show not only that this is a solution to the VSFT
equations of motion, but that we can define infinite many independent 
such solutions. We call such solutions {\it dressed slivers}.

These new solutions turn out to have an ill-defined $bpz$--norm
and action. To remedy this, we multiply the projector $P$ by a real 
parameter $\epsilon$,
thus creating an interpolating family of states between the sliver
($\epsilon=0$) and the dressed sliver ($\epsilon=1$). We define a
suitable regularization procedure for the determinants that appear
in this kind of trade,  by means of the
level truncation parameter $L$. Then we tune the latter to 
$1-\epsilon$ and  
show that the norm and the action corresponding to the dressed 
sliver can be made finite. A significant part of the paper is devoted
to the discussion of the regularization procedure. As it turns out,
there are many of them: by imposing the validity of the equation
of motion, we single out a hopefully consistent regularization scheme.

In a companion paper, \cite{BMP2}, the spectrum of states around 
the dressed sliver is 
analyzed. It will be seen there that this new type of solution solves 
several problems the spectrum around the sliver has.

The paper is organized as follows.
In section 2, after a brief summary  concerning the operatorial construction
of the sliver in the matter sector, we define the dressed sliver state by
using a ``half string vector'' $\xi$; we  show that such a state  solves the
zero momentum projector equations.  Section 3 is devoted to the concrete
construction of such a vector $\xi$: we show that there are infinite many
numerable consistent choices that solve the projector equation.  In section 4
we implement a deformation procedure, which exhibits many interesting
features: first of all such  deformation  turns out to define a continuous 
family of states which is closed under the $*$--product; we show that, in such
a family, the $*$--product acts in a commutative way and that  the dressed
sliver is  the identity element, inverse elements and null elements are shown
to exist: we show that we can consistently extract two abelian subalgebras
which are isomorphic via an inversion map, each one containing an identity
element and a  null element as projectors. Section 5 deals with the
computation of the action of the dressed sliver, we implement a regularization
 relating the vanishing of the determinants involving infinite level Neumann
coefficients to infinities which arise due to dressing. We show that it is
possible to extract a finite answer for the action by appropriately choosing 
a limiting procedure.  In section 6 we extend our results to the ghost part in
a straightforward way, and show that also the ghost part of the action can be
made finite.  In section 8 some other finite--action solutions are discussed,
in particular the ones related to parallel coincident and lower dimensional D--branes.  In section 9 
some relations with previous works are discussed and some concluding
remarks are presented. Appendix A contains well known properties of matter and
ghost star algebra. Appendix B contains detailed calculations of the
determinants needed in the main text. Appendix C deals with subtleties 
of the limiting procedures which arise while regularizing the
solutions. Appendix D contains an
explicit proof concerning the $*$--product of dressed states in the ghost sector.

\section{Dressing the sliver}

To start with we recall some formulas relevant to VSFT. The action is
\beq
{\cal S}(\Psi)= - \frac 1{g_0^2} \left(\frac 12 \langle\Psi |{\cal Q}|\Psi\rangle +
\frac 13 \langle\Psi |\Psi *\Psi\rangle\right)\label{sftaction}
\eeq
where
\beq
{\cal {Q}} =  c_0 + \sum_{n>0} \,(-1)^n \,(c_{2n}+ c_{-2n})\label{calQ}
\eeq
Notice that the action (\ref{sftaction}) does not contain any 
singular normalization constant, as opposed to \cite{GRSZ2,HKw}. 
This important issue will be discussed in section 8. 
The equation of motion is 
\beq
{\cal Q} \Psi = - \Psi * \Psi\label{EOM}
\eeq
and the ansatz for nonperturbative solutions is in the factorized form
\beq
\Psi= \Psi_m \otimes \Psi_g\label{ans}
\eeq
where $\Psi_g$ and $\Psi_m$ depend purely on ghost and matter
degrees of freedom, respectively. Then eq.(\ref{EOM}) splits into
\be
{\cal Q} \Psi_g & = & - \Psi_g *_g \Psi_g\label{EOMg}\\
\Psi_m & = & \Psi_m *_m \Psi_m\label{EOMm}
\ee
where $*_g$ and $*_m$ refers to the star product involving only the ghost
and matter part.\\
The action for this type of solution becomes
\beq
{\cal S}(\Psi)= - \frac 1{6 g_0^2} \langle \Psi_g |{\cal Q}|\Psi_g\rangle
\langle \Psi_m |\Psi_m\rangle \label{actionsliver}
\eeq
$\langle \Psi_m |\Psi_m\rangle$ is the ordinary inner product, 
$\langle \Psi_m |$ being the $bpz$ conjugate of $|\Psi_m\rangle$ 
(see below).

We shall see later on how to find solutions to (\ref{EOMg}). For the time
being, as an introduction to the problem, let us concentrate on the 
matter part, eq.(\ref{EOMm}). The $*_m$ product is defined as follows
\beq
_{123}\!\langle V_3|\Psi_1\rangle_1 |\Psi_2\rangle_2 =_3\!\langle \Psi_1*_m\Psi_2|
\label{starm}
\eeq
where the three strings vertex $V_3$ is  
\beq
|V_3\rangle_{123}= \int d^{26}p_{(1)}d^{26}p_{(2)}d^{26}p_{(3)}
\delta^{26}(p_{(1)}+p_{(2)}+p_{(3)})\,{\rm exp}(-E)\,
|0,p\rangle_{123}\label{V3}
\eeq
with
\beq
E= \sum_{a,b=1}^3\left(\frac 12 \sum_{m,n\geq 1}\eta_{\mu\nu}
a_m^{(a)\mu\dagger}V_{mn}^{ab}
a_n^{(b)\nu\dagger} + \sum_{n\geq 1}\eta_{\mu\nu}p_{(a)}^{\mu}
V_{0n}^{ab}
a_n^{(b)\nu\dagger} +\frac 12 \eta_{\mu\nu}p_{(a)}^{\mu}V_{00}^{ab}
p_{(b)}^\nu\right) \label{E}
\eeq
Summation over the Lorentz indices $\mu,\nu=0,\ldots,25$
is understood and $\eta$ denotes the flat Lorentz metric.
The operators $ a_m^{(a)\mu},a_m^{(a)\mu\dagger}$ denote the non--zero
modes matter oscillators of the $a$--th string, which satisfy
\beq
[a_m^{(a)\mu},a_n^{(b)\nu\dagger}]=
\eta^{\mu\nu}\delta_{mn}\delta^{ab},
\quad\quad m,n\geq 1 \label{CCR}
\eeq
$p_{(r)}$ is the momentum of the $a$--th string and
$|0,p\rangle_{123}\equiv |p_{(1)}\rangle\otimes
|p_{(2)}\rangle\otimes |p_{(3)}\rangle$ is
the tensor product of the Fock vacuum
states relative to the three strings with definite c.m.  momentum . $|p_{(a)}\rangle$ is
annihilated by
the annihilation
operators $a_m^{(a)\mu}$ ($m\geq1$) and it is eigenstate of the momentum operator
$\hat p_{(a)}^\mu$
with eigenvalue $p_{(a)}^\mu$. The normalization is
\beq
\langle p_{(a)}|\, p'_{(b)}\rangle = \delta_{ab}\delta^{26}(p+p')\0
\eeq
The symbols $V_{nm}^{ab},V_{0m}^{ab},V_{00}^{ab}$ will denote
the coefficients computed in \cite{GJ1, GJ2, Ohta, tope, leclair1, leclair2}.
We will use them in the notation of Appendix A and B of \cite{RSZ2}. 

To complete the definition of the $*_m$ product we must specify the 
$bpz$ conjugation properties of the oscillators
\beq
bpz(a_n^{(a)\mu}) = (-1)^{n+1} a_{-n}^{(a)\mu}\0
\eeq

In this paper we will mostly discuss solutions representing D25--branes 
(see a comment on lower dimensional branes at the end),
which are translationally invariant. As a consequence we set all the momenta
to zero. So the integration over the momenta will be dropped
and the only surviving part in $E$ will be the one involving $V_{nm}^{ab}$.
This is what we understand in the following by $*_m$, unless otherwise
specified.

Let us now return to eq.(\ref{EOMm}).  Its solutions are projectors of
the $*_m$ algebra. We recall the simplest one, the sliver. It is
defined by
\beq
|\Xi\rangle = \N e^{-\frac 12 a^\dagger Sa^\dagger}|0\rangle,\quad\quad
a^\dagger S a^\dagger = \sum_{n,m=1}^\infty a_n^{\mu\dagger} S_{nm}
 a_m^{\nu\dagger}\eta_{\mu\nu}\label{Xi}
\eeq 
This state satisfies eq.(\ref{EOMm}) provided the matrix $S$ satisfies
the equation
\beq
S= V^{11} +(V^{12},V^{21})(1-
\Sigma{\cal V})^{-1}\Sigma
\left(\matrix{V^{21}\cr V^{12}}\right)\label{SS}
\eeq
where
\beq
\Sigma= \left(\matrix{S&0\cr 0& S}\right),
\quad\quad\quad
{\cal V} = \left(\matrix{V^{11}&V^{12}\cr V^{21}&V^{22}}\right),
\label{SigmaV}
\eeq
The proof of this fact is well--known. First one expresses 
eq.(\ref{SigmaV}) in terms of the twisted matrices $X=CV^{11},X_+=CV^{12}$
and $X_-=CV^{21}$, together with $T=CS=SC$, where 
$C_{nm}= (-1)^n\delta_{nm}$. The matrices $X,X_+,X_-$ are mutually 
commuting. Then, requiring that $T$ commute with them as well, one can show 
that eq.(\ref{SigmaV}) reduces to the algebraic equation
\beq
XT^2-(1+X)T+X=0\label{algeq}
\eeq
The interesting solution is
\beq
T= \frac 1{2X} (1+X-\sqrt{(1+3X)(1-X)})\label{sliver}
\eeq

The normalization constant $\N$ is calculated to be
\beq
\N= (\Det (1-\Sigma \V))^{\frac{D}{2}}\label{norm}
\eeq
where $D=26$.
The contribution of the sliver to the matter part of the action 
(see (\ref{actionsliver})) is given by
\beq
\langle \Xi|\Xi\rangle = \frac {\N^2}{(\det (1-S^2))^{\frac{D}{2}}}
\label{ener}
\eeq
Both eq.(\ref{norm}) and (\ref{ener}) are ill--defined and need to be 
regularized, after which they both turn out to vanish. This subject
will be taken up again in section 5.

In Appendix A we collect a series of properties and results concerning
the matrices $X,X_-,X_+,T$, together with other formulas that will be
needed in the following.

Now we want to deform the sliver by adding some special matrix to $S$. 
To this end  we first introduce the infinite vector $\xi=\{\xi_{n}\}$  
which is  chosen to satisfy the condition
\beq
\rho_1 \xi =0,\quad\quad \rho_2 \xi =\xi, \label{xi1}
\eeq
Notice that this vector does not have any Lorentz label 
(compare with \cite{RSZ3}). Next we set
\beq
\xi^T \frac 1{1-T^2}\xi =1 ,\quad\quad
\xi^T \frac {T}{1-T^2}\xi= \k
\label{noncond}
\eeq
where $^T$ denotes matrix transposition. Eqs.(\ref{noncond})
will be studied in section 3.
Our candidate for the {\it dressed sliver} solution is given by an
ansatz similar to (\ref{Xi}) 
\beq
|\hat\Xi\rangle = \hat{\cal N} e^{-\frac 12 a^\dagger \hat S a^\dagger}\ket 0,
\label{Xihat}
\eeq
with $S$ replaced by
\beq
\hat S = S +R,\quad\quad R_{nm}= \frac 1{\kappa +1}\left(\xi_n(-1)^m\xi_m
+\xi_m(-1)^n\xi_n\right)\label{Shat}
\eeq
As a consequence $T$ is replaced by
\beq
\hat T = T +P,\quad\quad  P_{nm}=
\frac 1{\kappa +1}\left(\xi_m\xi_n+\xi_n(-1)^{m+n}\xi_m\right) \label{That}
\eeq
From time to time a bra and ket notation will be used to represent $P$:
\beq
P=\frac 1{\kappa +1}
\left(|\xi\rangle\langle\xi|+|C\xi\rangle\langle C\xi|
\right)\label{P}
\eeq
We require the dressed sliver to satisfy hermiticity, which amounts to 
imposing that the $bpz$--conjugate state coincide with the hermitean 
conjugate one. This in turn implies
\be
|\xi\rangle\langle C\xi|+|\xi\rangle\langle C\xi|=
|\xi^*\rangle\langle C\xi^*|+|\xi^*\rangle\langle C\xi^*|\0
\ee
We satisfy this condition by choosing $\xi$ real. This means that 
$\kappa$ is real (and negative). We remark at this point that
the conditions (\ref{noncond}) are not very stringent. The only thing
one has to worry is that the $lhs$'s are finite (this is the only true
condition). Once this is true the rest follows from suitably rescaling
$\xi$, so that the first equation is satisfied, and from the reality of 
$\xi$ (see also next section).

We claim that $|\hat\Xi\rangle$ is a projector.
The dressed sliver matrix $\hat T$ does not commute with $X,X_-,X_+$
(as T does), but we can nevertheless make use of the property
$C\hat T= \hat T C$, because $CP=PC$. 
To prove our claim we must show that 
\beq
V^{11} +(V^{12},V^{21})({1}-
\hat\Sigma{\cal V})^{-1}\Sigma
\left(\matrix{V^{21}\cr V^{12}}\right)=\hat S\label{hatShatS}
\eeq
where
\beq
\hat\Sigma= \left(\matrix{\hat S&0\cr 0& \hat S}\right)\label{hatSigma}
\eeq
We will in fact prove in detail that
\beq
X +(X_{+},X_{-})({1}-
\hat{\cal T}{\cal M})^{-1}\hat\T
\left(\matrix{X_{-}\cr X_{+}}\right)=\hat T\label{hatTT}
\eeq
where
\be \label{MCNu}
\hat{\cal T}=C\hat\Sigma=\T+\CP,\quad\quad {\cal M}=C {\cal V}\0
\ee
To this end, let us define 
\beq
\hat\K = 1-\hat \T \M = 1-\T\M-\CP\M=\K-\CP\M\label{calK}
\eeq
The symbol $\K$ is the same as $\hat\K$ when the deformation $P$ is absent,
so it is the quantity relevant to the sliver.
Now we write
\be
\hat\K^{-1} = (1-\hat \T \M)^{-1}= \K^{-1} (1-\CP\M\K^{-1})^{-1}\0
\ee
We have
\beq
(1-\CP\M\K^{-1})^{-1}\CP = \left(\matrix{1&\rho_1-\kappa\rho_2\cr
\rho_2-\kappa\rho_1&1\cr}\right)\CP\label{inverseK}
\eeq
This can be shown either by expanding the $lhs$ in power series or multiplying
this equation from the left by $1-\CP\M\K^{-1}$ and verifying that it is
an identity. To obtain this result one must use eq.(\ref{noncond}) and 
the formulas in Appendix A, from which in particular one can derive
\be
X_+ \xi = X(T-1)\xi,\quad\quad X_-\xi =(1-XT)\xi\0
\ee
Now we can evaluate the $lhs$ of eq.(\ref{hatTT})
\be
&&X + (X_+,X_-) (1 -\hat \T \M)^{-1} \hat \T 
\left(\matrix{X_-\cr X_+\cr}\right)\0\\
&&=X+ 
(X_+,X_-) \K^{-1}(1-\CP \M\K^{-1})^{-1} (\T+\CP)
\left(\matrix {X_-\cr X_+\cr}\right)\0\\
&&=X +(\rho_1,\rho_2) \T \left(\matrix {X_-\cr X_+\cr}\right)+
(\rho_1,\rho_2)  (1-\CP \M\K^{-1})^{-1}\CP 
\left(\matrix {X_-\cr X_+\cr}\right)  \0\\
&&~~~~~~~~~~+(\rho_1,\rho_2)  (1-\CP \M\K^{-1})^{-1}\CP \M\K^{-1} \T
\left(\matrix {X_-\cr X_+\cr}\right)\0
\ee
The first two terms in the $rhs$ are exactly $T$. Next one notices that
\be
(\rho_1,\rho_2)(1-\CP \M\K^{-1})^{-1}=(1,1)\0
\ee
Therefore
\be
&&X + (X_+,X_-)  \hat\K^{-1} \hat \T 
\left(\matrix{X_-\cr X_+\cr}\right)=
T+PX_- + PX_+ + (1,1) P
\left(\matrix {TX\rho_2 +TX_+\rho_1\cr TX_-\rho_2+TX\rho_1\cr}\right)=\0\\
&&=T +\frac 1{\kappa+1}\Big(|C\xi\rangle\langle C\xi| X(T-1) +
|\xi\rangle\langle\xi| (1-XT) + |C\xi\rangle\langle C\xi|(1-XT) +
|\xi\rangle\langle\xi| X(T-1)\Big)\0\\
&& ~~+ \frac 1{\kappa+1}\Big(|C\xi\rangle\langle C\xi|T(1-XT) +
|\xi\rangle\langle\xi|XT + |C\xi\rangle\langle C\xi|XT +
|\xi\rangle\langle\xi|T(1-XT)\Big)=\0\\
&&= T + \frac 1{\kappa+1}\left(|\xi\rangle\langle\xi|+|C\xi\rangle\langle 
C\xi|\right) = T+P = \hat T\label{proof}
\ee
In the passage to the last line we have used the identity $XT-X+T-XT^2=0$.
This completes the proof that $\hat\Xi$ is a solution to (\ref{EOMm}).

We remark that, due to the arbitrariness of $\xi$, the result we have 
obtained brings into the game an 
infinite family of solutions to the equations of motion\footnote{We believe
this multiplicity of solutions to correspond mostly to gauge degrees 
of freedom.}. We shall see later
that this result can be easily generalized. For the time being however
we are interested in studying the properties of these new solutions.

The normalization constant $\hat\N$ is given by (see appendix B)
\beq
\hat\N =\Det (1-\hat\Sigma \V)^{\frac{D}{2}}= \Det(1-\T \M)^{\frac{D}{2}} 
\Det(1-\CP \M K^{-1})^{\frac{D}{2}}= 
\Det(1-\T \M)^{\frac{D}{2}} \cdot \frac 1{(\kappa+1)^{D}}\label{Nhat}
\eeq
However, if one tries to compute the norm of this state (which corresponds
to the its contribution to the action), i.e. 
$\langle \hat \Xi|\hat \Xi\rangle$, one finds an indeterminate result
(as will be apparent from the calculation in section 5). It is evident that
we have to introduce a regulator in order to end up with a finite action.
Our idea is to introduce a numerical parameter $\epsilon$ in front
of $R$ in the definition of $\hat\Xi$. In this way we define new squeezed 
states $\hat\Xi_\epsilon$ characterized by the matrix 
$\hat S_\epsilon=S+\epsilon R$.
But, before we come to that, a discussion of some issues concerning the
vector $\xi$ is in order. 

 \section{A discussion of $\xi$}

In this section we will give a precise construction of the ``half string'' 
vector $\xi$. In so doing it is very convenient to use the continuous $k$ 
basis of the star algebra.\\
In \cite{RSZ5} it was shown that the Neumann coefficients $(X,X_+,X_-)$ can 
be simultaneusly put in a continuous diagonal form as follows
\be
X=\int_{-\infty}^{\infty}dk\,  x(k)\,  |k\rangle\langle k|, \quad\quad
X_{\pm}=\int_{-\infty}^{\infty}dk\,  x_{\pm}(k)\,  |k\rangle\langle k|
\ee
The eigenvalues are given by\footnote{We hope the reader 
should not confuse $k$ with $\k$.} 
\be
x(k)&=& -\frac{1}{1+2\cosh(\frac{\pi k}{2})}\0\\
x_{\pm}(k)&=&\frac{1+\cosh(\frac{\pi k}{2})\pm\sinh(\frac{\pi k}{2})}
{1+2\cosh(\frac{\pi k}{2})}\0
\ee
and the eigenvectors
\be
|k\rangle&=&\left( \frac 2k \sinh(\frac{\pi k}{2})\right)^{-\frac 12}
\sum_{n=1}^{\infty} v_n(k) |n\rangle\0\\
v_n(k)&=&\sqrt{n}\oint \frac{dz}{2\pi i} \frac{1}{z^{n+1}}\, 
\frac{1}{k}\left(1-e^{-k \tan^{-1}(z)}\right)\0
\ee
These eigenvectors are normalized by the condition, \cite{Oku2},
\be
\langle k|k'\rangle =\delta(k-k')\0
\ee
In this basis the sliver matrix $T$ takes the remarkably simple form
\be
T=-\int_{-\infty}^{\infty}dk\, e^{-\frac{\pi|k|}{2}}\,  
|k\rangle\langle k|\0
\ee
One should think at the real line spanned by $k$ as a parametrization 
of the string itself in which the midpoint is represented by the $k=0$ 
eigenvector and the left (right) half  by  $k>0$ ($k<0$).
This is easy to see once the form of the projectors $\rho_1,\rho_2$ is 
given in such a basis
\be\label{diagrho}
\rho_1&=&\int_{-\infty}^{\infty}dk\,  \theta(k)\,  
|k\rangle\langle k|=\int_{0}^{\infty}dk\,   |k\rangle\langle k|\\
\rho_2&=&\int_{-\infty}^{\infty}dk\,  \theta(-k)\,  
|k\rangle\langle k|=\int_{0}^{\infty}dk\,   |-k\rangle\langle -k|\0
\ee
The value of these projectors at $k=0$ is a subtle point \cite{HK}
and  we will 
avoid this singular mode in the constrution of the vector $\xi$.
Since  $\xi$ is constrained by $\rho_2|\xi\rangle=|\xi\rangle$, 
$\rho_1|\xi\rangle=0$, it is natural to parametrize it as
\be
|\xi\rangle=\int_{0}^{\infty}dk\,  \xi(y)\,  |-k\rangle
\ee
where $y=\frac{\pi k}{2}$.
Now the vector $|\xi\rangle$ is represented by the function $\xi(y)$, 
which has support on the positive real axis.
The expressions (\ref{noncond}) take the integral form
\be
\langle\xi|\frac{1}{1-T^2}|\xi\rangle&=&\frac{2}{\pi}\int_0^\infty dy\, 
 \xi^2(y) \frac{1}{1-e^{-2y}}=1\label{noncondint1}\\
\langle\xi|\frac{T}{1-T^2}|\xi\rangle&=&\frac{2}{\pi}\int_0^\infty dy\,  
\xi^2(y) \frac{-e^{-y}}{1-e^{-2y}}=\kappa\label{noncondint2}
\ee
Note that the denominator $1-T^2$ vanishes at $k=0$, so, in order 
to avoid infinities, we further require $\xi(y)$ to vanish
rapidly enough at $y=0$.
This means that the vector $\xi$ does not excite the (zero momentum)
midpoint mode.
The space of functions with support on the positive axis, vanishing at 
the origin, and satisfying (\ref{noncondint1}, \ref{noncondint2}) 
with finite $\kappa$, 
are spanned by 
a (numerable) infinite set of ``orthogonal'' functions defined by
\be
\xi_n(y)=\left(\frac{\pi}{2}(1-e^{-2y})e^{-y}\right)^{\frac 12} L_n(y)
\ee
where $L_n(y)$ is the $n$-th Laguerre polynomial.\\
The normalization factor in front of the polynomials has been chosen 
in order to satisfy
\be
\langle\xi_n|\frac{1}{1-T^2}|\xi_m\rangle=\int_0^\infty dy\, e^{-y} 
L_n(y)L_m(y)=\delta_{nm}
\ee
In a similar fashion, using standard properties of Laguerre 
polynomials\footnote{In particular we need the relation 
\be 
L_n(\lambda y)=\sum_{p=0}^n\left(\matrix{n\cr p}\right) 
\lambda^{n-p}(1-\lambda)^p L_{n-p}(y)\0
\ee }, one can prove that
\be
\langle\xi_n|\frac{T}{1-T^2}|\xi_m\rangle=-\int_0^\infty dy\, 
e^{-2y} L_n(y)L_m(y)=K_{nm}=
-\frac{1}{2^{n+m}}\frac{(m+n)!}{n!m!}
\ee
A simple numerical analysis shows that the eigenvalues of the matrix 
$K_{nm}$ lie in the interval $(-1,0)$. This is of course 
what one should expect once the normalization condition 
$\langle\xi\frac{1}{1-T^2}\xi\rangle=1$ is imposed. In fact the 
condition (\ref{noncondint2}) 
differs from (\ref{noncondint1}) by the insertion of the matrix $T$, 
which has a 
spectrum covering  (twice) the interval $(-1,0)$.\\
In order to prove that these half string vectors can be concretely 
defined as Fock space vectors,  we shall see that it is possible 
to have a complete control on their norm as well,  and that such 
norms are always positive . Using the same standard manipulations 
as before, we have
\be\label{normxi}
&&\langle\xi_n|\xi_m\rangle=
\langle\xi_n|\frac{1-T^2}{1-T^2}|\xi_m\rangle=\\
&&=\int_0^\infty dy\, e^{-y}(1-e^{-2y}) L_n(y)L_m(y)=\delta_{nm}-
\frac{2^{m-n}}{3^{n+m+1}}\sum_{p=0}^{n}4^p \  
\left(\matrix{n\cr p}\right)\left(\matrix{m\cr m-n+p}\right)\0
\ee
Again a simple numerical analysis shows that the eigenvalues 
of the matrix defined by the $rhs$ of (\ref{normxi}), lie in the
 interval $(0,1)$: this definitely ensures the existence of such 
vectors. As we will see in the last section, we can build orthogonal 
projectors (in the sense of the star product and of the $bpz$-norm)
by simply using 
different and orthogonal half-string vectors, where orthogonality is 
understood in the following sense
\be
\langle\xi|\frac{1}{1-T^2}|\xi'\rangle=0,\quad\quad 
\langle\xi|\frac{T}{1-T^2}|\xi'\rangle=0, 
\label{orth}
\ee
In view of the above discussion it is obvious that one  can always find 
a finite number of $\xi_n$'s to construct any given number of mutually 
orthogonal vectors although the 
number of $\xi_n$'s needed increases faster with respect to the number 
of orthogonal projectors.

\section{The states $\hat\Xi_\epsilon$}
 After the digression of the previous section, let us return to the problem
of regularizing the norm for the matter part of the dressed 
sliver. As anticipated in section 2, the (naive) definition (\ref{Xihat})
given in section 2 for the dressed sliver does not avoid ambiguities and 
indefiniteness, when
we come to compute its norm. The determinants involved in such calculations
are in general not well--defined. To evade this problem our idea is to deform
the dressed sliver by introducing a parameter $\e$, so that we get the dressed sliver when $\e=1$. When $\e\neq 1$ the state we obtain
is in general not a $*$--algebra projector.
We will define the norm of the dressed sliver as the limit of a sequence of
such states.

Let us introduce the state
\beq
|\hat\Xi_\e\rangle = \hat{\cal N_\e} e^{-\frac 12 a^\dagger \hat S_\e a^\dagger}|0\rangle,
\label{Xiehat}
\eeq
where
\beq
\hat S_\e= S +\e R, \label{Sehat}
\eeq
As a consequence $T$ is replaced by
\beq
\hat T_\e = T +\e P, \label{Tehat}
\eeq
The states defined in this way are not in general projectors, but have
very interesting properties. Not all of them are needed for the
following developments in this paper. However it is worth to make a short 
detour to illustrate them.

 We would like to show that the states (\ref{Xiehat})
define a continuous $*$--abelian 1--parameter family of states. 
First we show that they
are closed under the $*$--product.
Hence let us consider 
\be
|\hat\Xi_{\e_1}\rangle *|\hat\Xi_{\e_2}\rangle= 
\hat{\cal N}(\e_1,\e_2)e^{-\frac 12 a^\dagger C(T+\e_1P)*(T+\e_2P) 
a^\dagger}|0\rangle
\ee
where we denote
\beq
(T+\e_1P)*(T+\e_2P)\equiv X +(X_+,X_-)({1}-
\hat{\cal T}_{\e_1\e_2}{\cal M})^{-1}\hat{\cal T}_{\e_1\e_2}
\left(\matrix{X_-\cr X_+}\right)\label{hatT1T2}
\eeq
and
\beq
\hat{\cal T}_{\e_1\e_2}\equiv \left(\matrix{\hat T_{\e_1}&0\cr 0& 
\hat T_{\e_2}}\right)\label{hatT12}
\eeq
In order to compute this expression we need the generalized formula
\beq
(1-\CP_{\e_1\e_2}\M\K^{-1})^{-1}\CP_{\e_1\e_2} =\frac{1}{1+(1-\e_1)
(1-\e_2)\kappa}
\left(\matrix{(1-\e_2)\kappa+1&\e_1(\rho_1-\kappa\rho_2)\cr
\e_2(\rho_2-\kappa\rho_1)&(1-\e_1)\kappa+1\cr}\right)
\CP_{\e_1\e_2}\label{inverseK12}
\eeq
One can prove this formula as an easy generalization of section 2. 
Alternatively one can check it directly by multiplying it on the left with 
$(1-\CP_{\e_1\e_2}\M\K^{-1})$ (a detailed proof can be found in the 
appendix B). Then things are straightforward and we get
\be
(T+\e_1P)*(T+\e_2P)=T+(\e_1\star\e_2)P
\ee
where we have defined the {\it abelian} multiplication law between real 
numbers
\beq \label{estare}
\e_1\star\e_2=\frac{\e_1\e_2}{1+(1-\e_1)(1-\e_2)\kappa}
\eeq
This product is easily shown to be associative
\be
(\e_1\star\e_2)\star\e_3&=&\e_1\star(\e_2\star\e_3)=\label{2star}\\
&=&\frac{\e_1\e_2\e_3}{1+\kappa\left(2-\e_1-\e_2-\e_3+\e_1\e_2\e_3+
\kappa(1-\e_1)(1-\e_2)(1-\e_3)\right)}\0
\ee
and  exhibits three idempotent elements
\be
0\star 0=0,\quad\quad
1\star 1=1, \quad\quad
\frac{\k+1}{\k}\star\frac{\k+1}{\k}=\frac{\k+1}{\k}\0
\ee
Note that $1$ is the identity
\be
\e\star1=1\star\e=\e\0
\ee
The inverse with respect to this product is given by
\beq\label{inverse}
\e^{\star-1}=\frac{(1-\e)\k+1}{(1-\e)\k+\e}
\eeq
so that
\be
\e\star\frac{(1-\e)\k+1}{(1-\e)\k+\e}=1\0
\ee
We have two distinct null elements which are $0$ and  $\frac{\k+1}{\k}$
\be
0\star\e=\e\star0=0,\quad\quad
\frac{\k+1}{\k}\star\e=\e\star\frac{\k+1}{\k}=\frac{\k+1}{\k}\0
\ee
The point $\{\infty\}$ is naturally in the domain as it can be reached 
from any $\e\neq 0,1,\frac{\k+1}{\k}$ by $\star$--product
\be
\e\star\left(1+\frac{1}{(1-\e)\k}\right)=\infty\0
\ee
The simultaneus presence of two  null elements makes their product 
ambiguous 
\be
\frac{\k+1}{\k}\star 0={\rm indeterminate}\0
\ee
Note in particular that we have
\be
0^{\star-1}=\frac{\k+1}{\k},\quad\quad
\left(\frac{\k+1}{\k}\right)^{\star-1}=0\0
\ee
This is very reminescent of what happens with real numbers when they 
are completed with $\infty$, in which case what is ambiguous is the product 
$0\cdot\infty$. One should note actually that this is the same situation,  
deformed by the parameter $\k$, as in the limit $\k\rightarrow0$ one 
recovers the usual product and, in particular, 
$\frac{\k+1}{\k}\rightarrow\infty$.
                
In view of the structure we have found, two new abelian subalgebra of the 
$*$--product  are naturally identified. The first is   
$R\cup\{\infty\}\setminus\{\frac{\k+1}{\k}\}$ and  contains, as projectors, 
the sliver ($\e=0$) and the dressed 
sliver ($\e=1$). The second is 
$R\cup\{\infty\}\setminus\{0\}$ and contains the projectors $\e=1$ and 
$\e=\frac{\k+1}{\k}$. We will call the state identified by 
$\e=\frac{\k+1}{\k}$ the {\it exotic} dressed sliver. Note also that these 
two algebras are isomorphic to each other via the inversion map 
(\ref{inverse}).\\
Since we are dealing with projectors, normalization is needed. 
The normalization of all the states in the two algebras is completely 
fixed once we ask the sliver and the exotic sliver to be really null 
elements.
A general element of the two algebras can be written as
\beq
\ket{\hat\Xi_\e}^{(1,2)}={\cal N}_\e^{(1,2)}e^{-\frac{\e x}{\k+1}}\ket\Xi
\label{xi12}
\eeq
where $\ket\Xi$ is the usual sliver with its (vanishing) normalization 
and the superscript $^{(1,2)}$ labels the algebras, moreover we have 
identified 
\be
x=a^\dagger_\mu C (\ket\xi\bra\xi+\ket{C\xi}\bra{C\xi})
a^\dagger_\nu \eta^{\mu\nu}.\0
\ee
It is then easy to show that the star products of two such states is 
given by
\beq
\ket{\hat\Xi_\e}^{(1,2)}*\ket{\hat\Xi_\eta}^{(1,2)} =
\frac{{\cal N}_\e^{(1,2)}{\cal N}_\eta^{(1,2)}}{{\cal N}_{\e\star\eta}^{(1,2)}}
\left(\frac{\k+1}{1+(1-\e)(1-\eta)\k}\right)^{\!D}
\ket{\hat\Xi_{\e\star\eta}}^{(1,2)}\label{mspeet}
\eeq
The second factor in the $rhs$ comes from 
$\Det(1-\hat\T_{\e\eta}\M)^{-\frac 12}$, 
see appendix B.
In the first algebra the null element is the sliver $(\e=0)$ and, of 
course, ${\cal N}_0^{(1)}=1$  since the sliver is a projector by itself. 
The star product with another state of the same algebra is then
\be
\ket{\hat\Xi_0}  ^{(1)}*\ket{\hat\Xi_\e}^{(1)}={\cal N}_\e^{(1)}
\left(\frac{\k+1}{1+(1-\e)\k}\right)^{\!D}\ket{\hat\Xi_0}^{(1)}\0
\ee
which implies
\beq
{\cal N}_\e^{(1)}=\left(\frac{1+(1-\e)\k}{\k+1}\right)^{\!D}\label{norm1}
\eeq
With this  choice of normalization we have, use eq.({\ref{estare}),
\be
\frac{{\cal N}_\e^{(1)}{\cal N}_\eta^{(1)}}{{\cal N}_{\e\star\eta}^{(1)}}=
\left(\frac{1+(1-\e)(1-\eta)\k}{\k+1}\right)^{\!D}\0
\ee
so the first algebra closes with structure constant 1,
\beq
\ket{\hat\Xi_\e}^{(1)}*\ket{\hat\Xi_\eta}^{(1)}=
\ket{\hat\Xi_{\e\star\eta}}^{(1)}\label{nostruc}
\eeq
Note that the exotic sliver has, in this algebra, an extra vanishing 
normalization due to the dressing factor, (\ref{norm1}),
so it is naturally excluded.
Concerning the inverse algebra one has first to note that, in order for 
the exotic sliver to be a projector it should be that
\be
{\cal N}_\frac{\k+1}{\k}^{(2)}=\frac{1}{\k^D}\0
\ee
Now one should ask the exotic sliver to be a null element of the algebra
\be
\ket{\hat\Xi_{\frac{\k+1}{\k}}}^{(2)}*\ket{\hat\Xi_\e}^{(2)}=\ket{\hat\Xi_{\frac{\k+1}{\k}}}^{(2)}
\0
\ee
which implies
\beq
{\cal N}_\e^{(2)}=\left(\frac{\e}{\k+1}\right)^{\!D}\label{norm2}
\eeq
In this case too the inverse algebra closes with structure constant 1,
\beq
\ket{\hat\Xi_\e}^{(2)}*\ket{\hat\Xi_\eta}^{(2)}=\ket{\hat\Xi_{\e\star\eta}}^{(2)}
\label{nostr2}
\eeq
Note that the dressed sliver has the same normalization  and behaves as
the identity in both algebras.\\
The next task is to compute the $bpz$--norm of such states; here
we limit ourselves to a formal expression, since all of them 
are constructed on the sliver which is known to have vanishing norm.
This formal expression will be suitably regularized in the next section.
 
Using results from the appendix B, for states belonging to the $(1)$ 
algebra we obtain
\be
&&^{(1)}\!\bra{\hat\Xi_\e}\hat\Xi_\e\rangle^{(1)} = 
\left(\frac{({\cal N}_\e^{(1)})^2}{\det(1-\hat T_\e^2)^{\frac 12}}
\Det(1-\T\M)\right)^{\!D} \bra00\rangle\label{formalnorm} \\
&& = \frac{V^{(D)}}{(2\pi)^D} 
\left(\frac{(1+(1-\e)\k)^2}{(1-\e)(1+\k)(1+\k-\e(\k-1))}\right)^{\!D}
\left(\frac{\Det(1-\T\M)}{(\det(1-T^2))^{\frac12}}\right)^{\!D} 
\ee
As we have just mentioned, this expression is formal, since, due to the fact that the third factor 
in the $rhs$ is vanishing, all 
norms in this algebra vanish as well, except perhaps for $\e=1$ and 
$\e=\frac{\k+1}{\k-1}$, for which the denominator of the second factor 
vanishes, and we get a $\frac00$ expression.  

A remark is in order 
concerning the 
state represented by $\e=\frac{\k+1}{\k-1}$. This state is not a projector, 
but has the nice property of squaring to the dressed sliver, and 
can be therefore identified with a non trivial  ``square root'' of unity
\be
\ket{\hat\Xi_{\frac{\k+1}{\k-1}}}^{(1)}*\ket
{\hat\Xi_{\frac{\k+1}{\k-1}}}^{(1)}=\ket{\hat\Xi_1}^{(1)}\0
\ee
It is quite natural therefore that, if the dressed sliver can  have a 
finite norm, also its square root should.
For what concerns the inverse algebra $(2)$ all can be repeated with only 
slight modifications for the normalization factors ${\cal N}^{(2)}$, which
never vanishes for states belonging to the algebra itself. Again we can 
have a non--vanishing norm for the dressed sliver and its square root, 
which 
has the same normalization as in the algebra $(1)$. Therefore we will not 
repeat the computation of the norm. In any case, in the rest of the paper, 
we will deal only with the {\it first} algebra.

\section{The dressed sliver action: matter part.}
In the previous section we have introduced a Fock space state, depending
on a parameter $\epsilon$, that interpolates between the sliver 
$\epsilon=0$ and the dressed sliver $\epsilon=1$. Now we intent to show
that by its means, we can give a precise 
definition of the norm of the dressed sliver, so that both its norm and 
its action can be made finite.

As already mentioned above, the determinants in (\ref{norm}),
(\ref{ener}) relevant to the sliver are ill--defined. They are actually
well defined for any finite truncation of the matrix $X$ to level $L$ and 
need a regulator to account for its behaviour when $L\to\infty$.
A regularization that fits particularly our needs  was introduced by
Okuyama \cite{Oku2} and we will use it here. It consists in using an asymptotic
expression for the eigenvalue density $\rho(k)$ of $X$ (see also section 3),
$\rho(k) \sim \frac 1{2\pi} \log L+\rho_{fin}(k)$, for large $L$, where
$\rho_{fin}(k)$ is a finite contribution when $L\rightarrow\infty$, see
\cite{belov}. This leads to asymptotic expressions for
the various determinants we need. In particular the scale of $L$ can be chosen
in such a way that  \be
&&\det(1+T)= h_+\,L^{-\frac 13}+\ldots\0\\
&&\det(1-T)= h_-\,L^{\frac 16}+\ldots\label{asymp}\\
&&\det(1-X)= h_X\,L^{\frac 19}+\ldots\0
\ee
where dots denote non--leading 
contribution when $L\to\infty$ and $h_+,h_-,h_X$ are suitable numerical 
constants which arise due to the finite contribution in the eigenvalue
density\footnote{In particular, for any infinite matrix $A$ which is diagonal
in the $k$--basis, the determinant can be regularized by the level $L$ as 
\be
\det(A)=h_A\, L^{\int_{-\infty}^{\infty}\frac{dk}{2\pi}A(k)},\quad\quad
h_A=e^{\int_{-\infty}^{\infty}dk \rho_{fin}(k)A(k)}.\0
\ee
We thank D.Belov for a discussion on this point.}.
Our strategy consists in tuning $L$ with $\epsilon$ in such a way as to 
obtain finite results. Let us start, as a warm up exercise, with the {\it bpz} norm 
$\langle \hat \Xi_\epsilon |\hat \Xi_\epsilon \rangle$. We have
\beq
\langle \hat \Xi_\epsilon |\hat \Xi_\epsilon \rangle = 
\frac{\hat \N_\epsilon^2}{[\det(1-\hat S_\epsilon^2)]^{D/2}}\bra0|0\rangle
\label{actionsliver1}
\eeq
where $D=26$ and (see previous section)
\beq
\hat \N_\e = [\Det(1- \Sigma \V)]^{D/2} \N_\e^{(1)},\quad\quad 
\N_\e^{(1)}= \left(\frac {1+(1-\e)\kappa}{\k+1}\right)^{\!D}
\label{detSeV}
\eeq
 
Likewise we have
\be
\det(1-\hat S_\epsilon^2) &=& \det(1-\hat T_\epsilon^2) = 
\det(1-\hat T_\epsilon) 
\det(1+\hat T_\epsilon)\0\\
&=& \det(1-T^2) \det(1-\epsilon P \frac 1{1-T}) 
\det(1+\epsilon P \frac 1{1+T}) 
\label{hatS2}
\ee
Using the results of appendix B.3 we find
\beq \label{hatS2'}
\det(1-\hat S_\epsilon^2) = \det(1-T^2) (1-\epsilon)^2 
\left(\frac {\kappa+1-\epsilon(\kappa -1)}{\kappa+1}\right)^2
\eeq
Therefore, in the limit $\epsilon\to 1$ the dominant term will be
\beq
\det(1-\hat S_\epsilon^2) = \det(1-T^2) (1-\epsilon)^2 \left(\frac {2}
{\kappa+1}\right)^2\label{hatS2''}
\eeq
Now, recalling that $\Det(1- \Sigma\V) = \det (1-X)\det (1+T)$, and putting
together all the above results, we find
\beq \label{lime1}
\langle \hat \Xi_\epsilon |\hat \Xi_\epsilon \rangle = 
\left(h\,\frac{1}{4(\k+1)^2} 
\frac{L^{-\frac 5{18}}}{(1-\epsilon)^2}+\ldots\right)^{\!\frac D2}
\langle 0|0\rangle,\quad\quad h=\frac{h_X^2h_+}{h_-}
\eeq
where dots denote irrelevant terms in the limit $\epsilon\to 1$ and
$L\to \infty$.
Therefore, if we assume that 
\beq
1-\epsilon = s L^{-\frac 5{36}}\label{regpres}
\eeq
for some constant $s$, we have
\beq
\lim_{\e\to 1}\,\langle \hat \Xi_\epsilon |\hat \Xi_\epsilon \rangle 
=\left( \frac{h}{4(\k+1)^2 s^2} \right)^{\!\frac D2}
\langle 0|0\rangle
\label{finitenorm}
\eeq
which may take any prescribed positive finite value \footnote{It is obvious 
that the constants $\k+1$ and $h$ could be absorbed in $s$.} .
The factor $\langle 0|0\rangle= \delta^{(D)}(0)$ is normalized to
$\frac {V^{(D)}}{(2\pi)^D}$.
We notice for later use that, in order for such 
prescription to be consistent, it must be that if we rescale $1-\e$, 
$L^{-\frac 5{36}}$ should be accordingly rescaled so that their ratio is
always $s$. This is in order to guarantee that the limit be scale 
independent.  

It would look natural to define the number (\ref{finitenorm}) as the norm 
$\langle \hat \Xi|\hat \Xi\rangle$ of our {\it regularized dressed 
sliver}. However, as we shall see next, the regularization
prescription defined by eqs.(\ref{lime1},\ref{regpres},\ref{finitenorm})
does not guarantee that the equations of motion be satisfied in the action.
In fact, as it turns out (see below),
\beq
\lim_{\e\to 1} \,\langle \hat \Xi_\epsilon |\hat \Xi_\epsilon \rangle 
\neq \lim_{\e\to 1} \,\langle \hat \Xi_\epsilon 
|\Xi_\e*\hat \Xi_\epsilon \rangle \label{noeom}
\eeq

There is here a subtle problem. We delve into it by 
analyzing the quantity 
\beq
\langle \hat \Xi_{\e_1} |\hat \Xi_{\e_2} \rangle = 
\frac {\hat \N_{\e_1}
\hat \N_{\e_2}}{\det(1-\hat S_{\e_1}\hat S_{\e_2})}\langle 0|0\rangle 
\label{X1X2}
\eeq
The analysis carried out in Appendix C leads us to 
infinite many ways of taking the limit $\e_1,\e_2\to 1$, with results that 
vary in a finite range. At one extreme we have the result obtained
above, which corresponds to $\e_1=\e_2 =\e$. At the 
the other extreme we have the ordered limit 
\beq
\lim_{\e_1\to 1}\left(\lim_{\e_2\to 1} 
\langle \hat \Xi_{\e_1} |\hat \Xi_{\e_2} \rangle\right)\label{limlim}
\eeq
According to Appendix C, when $\e_1$ and $\e_2$ are in the vicinity of 1
we have
\be
&&\frac 1{\langle 0|0\rangle}\,\langle \hat \Xi_{\e_1} |\hat \Xi_{\e_2} 
\rangle\0\\
&& = 
\left(\frac {\Det(1-\Sigma\V)}{\sqrt{\det(1-S^2)}}\right)^D
\left(\frac 1{4(\k+1)^2}\right)^{\frac D2}
\left(\frac{4}{(\k(1-\e_1)(1-\e_2)+1-\e_1\e_2)^2}\right)^{\frac D2}+
\ldots\0
\ee
where dots denote non--leading terms. Now let as take the limit 
(\ref{limlim})
\be
&&\frac 1{\langle 0|0\rangle}\,\lim_{\e_1\to 1}\left(\lim_{\e_2\to 1} 
\langle \hat \Xi_{\e_1} |\hat \Xi_{\e_2}\rangle\right) \label{xi1xi2f}\\
&&= \lim_{\e_1\to 1} \left(\frac {\Det(1-\Sigma\V)}
{\sqrt{\det(1-S^2)}}\right)^D
\left(\frac 1{4(\k+1)^2}\right)^{\frac D2}
\left(\frac{4}{(1-\e_1)^2}\right)^{\frac D2}+\ldots\0\\
&&= \lim_{\e_1\to 1} 
\left(\frac{h}{(\k+1)^2 }\right)^{\!\frac D2}
\left(\frac {L^{-\frac 5{36}}}{1-\e_1}\right)^{\!D}+\ldots = 
\left(\frac{h}{(\k+1)^2 s_1^2}\right)^{\!\frac D2}\0
\ee
provided 
\beq
1-\e_1 =s_1 L^{-\frac 5{36}}\label{regpres1}
\eeq
It is easy to see that if we reverse the order of the limits in
(\ref{limlim}) we obtain the same result.

Between this result and (\ref{finitenorm}) there is a
discrepancy, a factor of 4. This factor can be absorbed into a
redefinition of $s$, $s_1=2s$. Had we adopted still another method of 
taking the limit we would have obtained a result in between. 
In conclusion there are infinite many ways
of deriving the norm starting from
$\langle \hat \Xi_{\e_1} |\hat \Xi_{\e_2} \rangle$, 
they all lead to the same result up to a redefinition of the $s$ factor.

Now the question is: do we have a criterion to select among 
all these different limits? The answer is: yes, we do.
It is the requirement that the equation of motion be satisfied,
i.e. we must have
\beq\label{weakeom}
\lim_{\e_1,\e_2\to 1} 
\langle \hat \Xi_{\e_1} |\hat \Xi_{\e_2}\rangle=
\lim_{\e_1,\e_2,\e_3\to 1}
\langle \hat \Xi_{\e_1} |\hat \Xi_{\e_2}*\hat \Xi_{\e_3}\rangle
\eeq
The analysis carried out in Appendix C of the expression in the $rhs$
tells us that once again there are infinite many ways to calculate the
triple limit, and there are infinite many ways to satisfy (\ref{weakeom}).
For instance, the limit $\e_1=\e_2=\e_3\to 1$ does not satisfy
(\ref{weakeom}), while the criterion of the ordered limits does, i.e.
that
\beq\label{boxlim}
\lim_{\e_1\to 1}\left(\lim_{\e_2\to 1} 
\langle \hat \Xi_{\e_1} |\hat \Xi_{\e_2}\rangle\right)=
\lim_{\e_1\to 1}\left(\lim_{\e_2\to 1}\left(\lim_{\e_3\to 1}
\langle \hat \Xi_{\e_1} |\hat \Xi_{\e_2}*\hat \Xi_{\e_3}\rangle\right)\right)
\eeq
First we notice that due to the symmetry of 
$\langle \hat \Xi_{\e_1} |\hat \Xi_{\e_2}*\hat \Xi_{\e_3}\rangle$ (see
Appendix C2),
the order $1,2,3$ in the last limit is irrelevant. What is relevant
is that the limits are taken in succession.
Now, using the formulas of the previous section and of Appendix B,
it is easy to see that
\beq
\lim_{\e_3\to 1}
\langle \hat \Xi_{\e_1} |\hat \Xi_{\e_2}*\hat \Xi_{\e_3}\rangle =
\lim_{\e_3\to 1}
\langle \hat \Xi_{\e_1} |\hat \Xi_{\e_2\star\e_3}\rangle
= \langle \hat \Xi_{\e_1} |\hat \Xi_{\e_2}\rangle\label{lll}
\eeq
Therefore (\ref{weakeom}) follows.

As we mentioned before, there are other ways of taking the limit
$\e_i\to 1$ while satisfying (\ref{weakeom}). However the 
ordered limits seem to have a privileged status, as we will try 
to explain next. We would like to show that the equation of motion 
(\ref{EOMm}) holds in a more general sense than eq.(\ref{boxlim}).
In other words we would like that
\beq\label{weaklim}
\lim_{\e_1\to 1} 
\langle \Psi |\hat \Xi_{\e_1}\rangle=
\lim_{\e_1,\e_2\to 1}
\langle \Psi |\hat \Xi_{\e_1}*\hat \Xi_{\e_2}\rangle
\eeq
for `any' state $\Psi$. 
In order to appreciate the problem one should recall that 
the limiting procedure is necessary whenever evaluations of determinants 
are involved, otherwise it is irrelevant. Therefore if $\Psi$ is any
state of the Fock space constructed by applying to the vacuum a 
polynomial of the string creation operators, eq.(\ref{weaklim}) holds;
the only proviso is that, since $\hat \Xi_\e$ contains a normalization
which vanishes when $L\to\infty$ (but it diverges in the ghost case,
see below), we must take this limit as the last operation.

The validity of eq.(\ref{weaklim}) may be in danger only when $\Psi$
is a close relative to $\hat \Xi$. We have already seen how to deal
with the case $\hat \Xi_\e$. The conclusion does not change
if the $\hat \Xi_\e$ is multiplied by a polynomial of the string 
creation operators or even by a coherent state contructed out of the 
latter. One may ask what happens when $\Psi$ coincides with $\hat \Xi$
itself. In this case the expressions under the limit symbols in
eq.(\ref{weaklim}) make sense, and we have to make sure that the
equation holds. It is easy to see that, once again, it holds with
the ordered limiting procedure. The set of states $\Psi$ for which
(\ref{weaklim}) holds, does not exhaust all the states one can think of,
however it contains all Fock space states as well as all the states 
that are relevant to this and the companion paper, \cite{BMP2}. To
characterize these limitiations we say that the EOM holds in a weak sense.
 
{\it From now on we assume the ordered limit procedure as the good limiting
procedure}. 
In particular the norm of $\hat \Xi$ is defined by eq.(\ref{xi1xi2f}).

What we have achieved so far is to prove that it is possible to assign
a finite positive number to the expression (norm) 
$\langle \hat \Xi|\hat \Xi\rangle$, in a way which is consistent with
the matter equation of motion. It does not mean that a state
exists in the Hilbert space which is the limit of $ \hat\Xi_\e$
when $\e\to 1$.
In order to show this one would have to prove that the number
$||\hat \Xi_{\epsilon_1} -\hat \Xi_{\epsilon_2}||^2 $ becomes
smaller and smaller when $\e_1,\e_2\to 1$. In such a case Hilbert space
completeness would guarantee the existence of a limiting state. 
Now
\beq
||\hat \Xi_{\epsilon_1} -\hat \Xi_{\epsilon_2}||^2 =
\langle \hat \Xi_{\e_1} |\hat \Xi_{\e_1}\rangle+
\langle \hat \Xi_{\e_2} |\hat \Xi_{\e_2}\rangle-
2\langle \hat \Xi_{\e_1} |\hat \Xi_{\e_2}\rangle\label{cauchy}
\eeq
The first two terms are similar to (\ref{actionsliver1}),
while the last term has been calculated in (\ref{xi1xi2}).
From the latter equation it is evident that the Cauchy condition
would be satisfied if the term in the expression in the second line 
of eq.(\ref{xi1xi2}) were to approach 1 when $\e_1,\e_2\to 1$. However,
as seen in Appendix C1, this quantity remains at a finite distance
from 1 when $\e_1,\e_2\to 1$, unless one takes $\e_1=\e_2$.
The conclusion is that we cannot satisfy the Cauchy 
condition for the sequence $\hat\Xi_\e$. 

Therefore, while the regularization procedure defined above 
guarantees that we can associate a positive finite number to
the symbol $\langle \hat \Xi|\hat \Xi\rangle$, it does not allow us
to associate any Hilbert space state to $\hat\Xi$.
{\it The state $\hat \Xi$ lives outside the Hilbert space.}
A careful treatment of this problem would require embedding the
string theory Hilbert space into a larger space with suitably defined
topology, according to which $\lim_{\e\to 1} \hat \Xi_\e= \hat \Xi$
makes full sense. This interesting issue goes beyond the scope of the 
present paper.

\section{The ghost dressed sliver}

In this section our purpose is to find the ghost companion of the
regularized dressed sliver solution discussed above.
The previous analysis for the matter part can be easily extended
also to the ghost part. 

Let us start with the  definition of the $*_g$ product:
\begin{equation}
|\widetilde{\Psi}\rangle *_g |\widetilde{\Phi}\rangle \, = \,
_1\langle\widetilde{\Psi}|\, _2\langle\widetilde{\Phi}
|\widetilde{V}_3\rangle\label{starghost}
\end{equation}
where the ghost part of the 3-strings vertex is defined by
\begin{equation} \label{gh3vert}
|\widetilde{V}_3\rangle = \exp\left[\sum_{r,s=1}^3 \left(
\sum_{n,m=1}^\infty 
c_n^{(r)\dagger} \widetilde{V}_{nm}^{rs} b_m^{(s)\dagger}
+ \sum_{n=1}^\infty c_n^{(r)\dagger} \widetilde{V}_{n0}^{rs} b_0^{(s)}
\right) \right]
\prod_{r=1}^{3} \left( c_0^{(r)}c_1^{(r)} \right) |0\rangle_{123}
\end{equation}
Here $c_n^{(r)}$ and $b_n^{(r)}$ are the standard ghost oscillator
modes of the $r$-th string, which satisfy
\be
\left\{b_n^{(r)},c_m^{(s)\dagger}\right\} = \delta_{nm}\delta_{rs} \;,
\qquad b_n^{(r)\dagger} = b_{-n}^{(r)} \;, 
\quad c_n^{(r)\dagger} = c_{-n}^{(r)}\0
\ee
and
$|0\rangle_{123}\equiv|0\rangle_1\otimes|0\rangle_2\otimes|0\rangle_3$ 
is the tensor product of the SL$(2,\mathbf{R})$-invariant ghost vacuum
states, normalized such that
\be
\langle0| c_1^\dagger c_0 c_1 |0\rangle = 1 \;.\0
\ee
The symbols $\widetilde{V}_{nm}^{rs}$ and $\widetilde{V}_{n0}^{rs}$
are coefficients computed in \cite{GJ2,Ohta,tope,Samu} and their properties
necessary for us here\footnote{For a more complete discussion see 
\cite{tope,MM}.} are listed in appendix A. The $bpz$ 
conjugation properties are
\be
bpz\left(c_{n}^{(r)}\right) = (-1)^{n+1}c_{-n}^{(r)} \;,\qquad
bpz\left(b_{n}^{(r)}\right) = (-1)^{n}b_{-n}^{(r)} \;.\0
\ee

It was shown in \cite{HKw} that there is a simple solution of the
ghost field equation (\ref{EOMg}) in the form of the squeezed state
\begin{equation}
|\widetilde{\Xi}\rangle = \widetilde{\N} \exp\left( 
\sum_{n,m=1}^\infty c_n^\dagger \widetilde{S}_{nm} b_m^\dagger \right)
c_1 |0\rangle \;,\label{ghostsl}
\end{equation}
where the matrix $\widetilde{S}$ satisfies the equation
\begin{equation} \label{ghsc}
\widetilde{S} = \widetilde{V}^{11} + 
(\widetilde{V}^{12},\widetilde{V}^{21})
(I-\widetilde{\Sigma}\widetilde{\mathcal{V}})^{-1} \widetilde{\Sigma}
\left(\! \begin{array}{c} \widetilde{V}^{21} \\ \widetilde{V}^{12}
\end{array}\! \right) \;,
\end{equation}
which has exactly the same form as (\ref{SS}) ($\widetilde{\Sigma}$ and 
$\widetilde{\mathcal{V}}$ are defined as in (\ref{SigmaV})), but now
with  tildes. As before one introduces
$\widetilde{X}=C\widetilde{V}^{11}$,
$\widetilde{X}_+=C\widetilde{V}^{12}$ and
$\widetilde{X}_-=C\widetilde{V}^{21}$ (see appendix A for properties).
As the $\widetilde{X}_i$'s satisfy the same algebraic relatons as the $X_i$'s,
one can construct solutions to (\ref{ghsc}) the same way as for the
matter part. The solution we are interested in, in terms of
$\widetilde{T}=C\widetilde{S}$, is
\be
\widetilde{T} = \frac{1}{2\widetilde{X}} \left( 1 + \widetilde{X} -
\sqrt{(1+3\widetilde{X})(1-\widetilde{X})}\, \right)\0
\ee
The normalization constant is  
\begin{equation} \label{ghsnorm}
\widetilde{\mathcal{N}} = 
-\left[\Det(1-\widetilde{\Sigma}\widetilde{\mathcal{V}})\right]^{-1}
\;.
\end{equation}
The contribution of the ghost part to the action is given by
\begin{equation} \label{ghsac}
\langle\widetilde{\Xi}|\mathcal{Q}|\widetilde{\Xi}\rangle = 
\widetilde{\N}^2 \det(1-\widetilde{S}^2)
\end{equation}
Now the determinants in eqs. (\ref{ghsnorm}) and (\ref{ghsac}) are both
vanishing, in such a way that the ghost part of the  action diverges (see below). 
When one combines this with the results for the matter part 
(using (\ref{ans}), (\ref{norm}) and (\ref{ener})) one finds 
\cite{Oku2} that both normalization constant and action of the overall
state vanish.

Following the analysis of the matter part, we consider deformations of
this solution. We introduce two real vectors 
$\beta=\{\beta_n\}$ and $\delta=\{\delta_n\}$ which
satisfy
\begin{equation}
\widetilde{\rho}_1 \beta
= \widetilde{\rho}_1 \delta = 0 ,\quad\quad \widetilde{\rho}_2 \beta=\beta,
\quad\quad  \widetilde{\rho}_2 \delta = \delta \;.\label{rhotilde}
\end{equation}
We also set 
\begin{equation}
\langle\beta| \frac{1}{1-\widetilde{T}^2} |\delta\rangle = 1 \;, 
\qquad
\langle\beta| \frac{\widetilde{T}}{1-\widetilde{T}^2} |\delta\rangle
= \widetilde{\kappa}
\end{equation}
where the first equation fixes the relative  normalization of $\beta$ and $\delta$, 
and the  second  defines  $\widetilde{\kappa}$. Note that one can repeat
the analysis of section 3: since the eigenvalues of the ghost sliver
matrix $\widetilde T$ are the opposite of the eigenvalues of the
corresponding matter matrix $T$, it follows that $\widetilde \k$ is
non--negative.
 
We now dress the ghost part of the sliver and introduce the squeezed state
\begin{equation} \label{ghdsl}
|\widehat{\widetilde{\Xi}}_{\tilde{\epsilon}}\rangle 
= \widehat{\widetilde{\N}}_{\tilde{\epsilon}} \,
e^{c^\dagger \widehat{\widetilde{S}}_{\tilde{\epsilon}} b^\dagger} c_1 |0\rangle
\end{equation}
where instead of $\widetilde{S}$ we now have
\be
\widehat{\widetilde{S}}_{\tilde{\epsilon}} = 
\widetilde{S} + \tilde{\epsilon}\widetilde{R} \;,
\qquad
\widetilde{R} = \frac{1}{\widetilde{\kappa}+1} 
 (|C\delta\rangle \langle\beta| + |\delta\rangle \langle C\beta|)\0
\ee
It it easy to see that 
$\widehat{\widetilde{S}}_{\tilde{\epsilon}}^*=C\widehat{\widetilde{S}}_{\tilde{\epsilon}}C$
for $\beta$, $\delta$ real, which means that the string field is real.

Let us now calculate 
$\widehat{\widetilde{\Xi}}_{\tilde{\epsilon}} *_g \widehat{\widetilde{\Xi}}_{\tilde{\eta}}$,
where both states have the same $\beta$ and $\delta$. If one defines,
the reduced $*_{b_0}$--product as,\cite{Oku1},
\begin{equation} \label{ghrsp}
\widehat{\widetilde{\Xi}}_{\tilde{\epsilon}} *_{b_0} 
\widehat{\widetilde{\Xi}}_{\tilde{\eta}} \equiv 
b_0 \left( \widetilde{\Xi}_{\tilde{\epsilon}} *_g
\widehat{\widetilde{\Xi}}_{\tilde{\eta}} \right) \;,
\end{equation}
then one can immediately see that it can be calculated using the vertex
(\ref{gh3vert}) but \emph{without} terms containing $b_0$ modes
(reduced vertex). Then
the calculation of the reduced product (\ref{ghrsp}) repeats
essentially the calculation in the matter sector of
sec.\ 4, the only differences being that untilded objects  
are replaced by the corresponding tilded ones and, more important,
the determinants are raised to the power -$2/D$ with respect
to the corresponding matter ones (this is because of the 
anticommutativity of ghosts). The result is then
\begin{eqnarray}
| \widehat{\widetilde{\Xi}}_{\tilde{\epsilon}} \rangle *_{b_0}
 | \widehat{\widetilde{\Xi}}_{\tilde{\eta}} \rangle &=& 
 \frac{\widehat{\widetilde{\mathcal{N}}}_{\tilde{\epsilon}} \,
 \widehat{\widetilde{\mathcal{N}}}_{\tilde{\eta}}}{
 \widehat{\widetilde{\mathcal{N}}}_{\tilde{\epsilon} \star \tilde{\eta}}} \,
 \Det(1-\widetilde{\Sigma}_{\tilde{\epsilon}\tilde{\eta}}
 \widetilde{\mathcal{V}}) \,
 |\widehat{\widetilde{\Xi}}_{\tilde{\epsilon} \star \tilde{\eta}}\rangle 
 \nonumber \\
&=& \frac{\widehat{\widetilde{\mathcal{N}}}_{\tilde{\epsilon}} \,
 \widehat{\widetilde{\mathcal{N}}}_{\tilde{\eta}}}{
 \widehat{\widetilde{\mathcal{N}}}_{\tilde{\epsilon} \star \tilde{\eta}}}
 \left[\frac{1+(1-\tilde{\epsilon})(1-\tilde{\eta})
 \widetilde{\kappa}}{\widetilde{\kappa}+1} \right]^2
 \Det(1-\widetilde{\Sigma}\widetilde{\mathcal{V}}) \,
 |\widehat{\widetilde{\Xi}}_{\tilde{\epsilon} \star \tilde{\eta}}\rangle \;,
\label{ghrpds}
\end{eqnarray}
where the $\star$ multiplication rule is defined in (\ref{estare}).

Now, it was shown in \cite{Oku1} that, if states $A$ and $B$ are in 
the subspace spanned
by coherent states, then the $*_g$ product can be obtained from the
reduced product using
\begin{equation}
A *_g B = \mathcal{Q} \left(A *_{b_0} B\right) \;,
\end{equation}
which applied to (\ref{ghrpds}) gives
\begin{equation} \label{ghspds}
| \widehat{\widetilde{\Xi}}_{\tilde{\epsilon}} \rangle *_g
| \widehat{\widetilde{\Xi}}_{\tilde{\eta}} \rangle =
\frac{\widehat{\widetilde{\mathcal{N}}}_{\tilde{\epsilon}} \,
\widehat{\widetilde{\mathcal{N}}}_{\tilde{\eta}}}{
\widehat{\widetilde{\mathcal{N}}}_{\tilde{\epsilon} \star \tilde{\eta}}}
\left[\frac{1+(1-\tilde{\epsilon})(1-\tilde{\eta})
\widetilde{\kappa}}{\widetilde{\kappa}+1} \right]^2
\Det(1-\widetilde{\Sigma}\widetilde{\mathcal{V}}) \mathcal{Q}
|\widehat{\widetilde{\Xi}}_{\tilde{\epsilon} \star \tilde{\eta}}\rangle \;.
\end{equation}

However, a more careful derivation is needed because states like (\ref{ghdsl}) are
(at least apparently) not of the required form and it could be risky
to use the above argument. In appendix D we give a direct 
proof that (\ref{ghspds}) is correct.

At this point one should observe a formal similarity between eq.
(\ref{ghspds}) and the corresponding one in the matter sector 
(\ref{mspeet}). In fact, one can now basically repeat the 
arguments of sections 4 and 5 with only minor modifications.

First, it is natural to choose a normalization such that (\ref{ghspds}) has
the following form
\begin{equation} \label{ghnspeet}
|\widehat{\widetilde{\Xi}}_{\tilde{\epsilon}}\rangle
*_g |\widehat{\widetilde{\Xi}}_{\tilde{\eta}}\rangle
= -\mathcal{Q} 
|\widehat{\widetilde{\Xi}}_{\tilde{\epsilon}\star\tilde{\eta}}\rangle
\end{equation}
Again, there are two different normalizations with this property, 
given by
\begin{eqnarray} \label{ghnorm1}
\widehat{\widetilde{\N}}_{\tilde{\epsilon}}^{(1)} &=& -\left(
 \frac{\widetilde{\kappa}+1}{1+(1-\tilde{\epsilon})\widetilde{\kappa}}
 \right)^{\!2}
 \left[\Det(1-\widetilde{\Sigma}\widetilde{\mathcal{V}})\right]^{-1}
 \\
\widehat{\widetilde{\N}}_{\tilde{\epsilon}}^{(2)} &=& -\left(
 \frac{\widetilde{\kappa}+1}{\tilde{\epsilon}}\right)^{\!2}
 \left[\Det(1-\widetilde{\Sigma}\widetilde{\mathcal{V}})\right]^{-1}
 \;.
\end{eqnarray}
The first one is singular in 
$\tilde{\epsilon}=\frac{\tilde{\kappa}+1}{\tilde{\kappa}}$, and the 
second one in $\tilde{\epsilon}=0$. From now on we use exclusively the
first one, and drop the $^{(1)}$ in superscript.

From (\ref{ghnspeet}) it follows that our states (\ref{ghdsl}) satisfy the
ghost equation of motion when
\begin{equation}
\tilde{\epsilon} \star \tilde{\epsilon} = \tilde{\epsilon}
\end{equation}
and we already know that it is true only for 
\begin{equation}
\tilde{\epsilon} = 0,\, 1,\, 
\frac{\widetilde{\kappa}+1}{\widetilde{\kappa}}
\end{equation}
Again, beside the Hata-Kawano solution (i.e. the solution
with $\tilde{\epsilon}=0$), we obtain in
addition two families of solutions, depending on the choice of $\beta$ and 
$\delta$.

Now we show that for the solution with $\tilde{\epsilon}\to1$ (ghost
part of the dressed sliver) we can define a finite action. We shall 
consider first the kinetic term, for which we need
\begin{eqnarray} 
\langle\widehat{\widetilde{\Xi}}_{\tilde{\epsilon}_1}|
 \mathcal{Q} \,| \widehat{\widetilde{\Xi}}_{\tilde{\epsilon}_2}\rangle &=&
 \langle\widehat{\widetilde{\Xi}}_{\tilde{\epsilon}_1}|
 c_0|\widehat{\widetilde{\Xi}}_{\tilde{\epsilon}_2}\rangle
 =  \widehat{\widetilde{\mathcal{N}}}_{\tilde{\epsilon}_1} \,
 \widehat{\widetilde{\mathcal{N}}}_{\tilde{\epsilon}_2} \,
 \det(1 - \widehat{\widetilde{S}}_{\tilde{\epsilon}_1}
 \widehat{\widetilde{S}}_{\tilde{\epsilon}_2}) \nonumber \\
&=& \left( 1 - \prod_{i=1}^2
 \frac{\tilde{\epsilon}_i}{1+(1-\tilde{\epsilon}_i)\tilde{\kappa}}
 \right)^{\!2}
 \frac{\det(1 - \widehat{\widetilde{S}}^2)}{\left[
 \Det(1 - \widetilde{\Sigma} \widetilde{\mathcal{V}})\right]^2} \;.
\label{ghace}
\end{eqnarray}
where in the last line we used (\ref{ghnorm1}) and (\ref{1-t1t2}). It
was shown in \cite{Oku2} that the level truncation regularization 
at the leading order leads to
\begin{equation} \label{ghokl}
\frac{\det(1 - \widehat{\widetilde{S}}^2)}{\left[
\Det(1 - \widetilde{\Sigma} \widetilde{\mathcal{V}})\right]^2}
= \frac{L^{\frac{11}{18}}}{\tilde h} + \ldots
\end{equation}
where $\tilde h$ is the numerical factor analogous to $h$ for the ghost part.
The $rhs$ of (\ref{ghokl}) diverges when the cutoff is lifted, i.e., when $L\to\infty$. But, as
for the matter part, we see that if we let $\tilde{\epsilon}\to1$
in a specific way, the expression (\ref{ghace}) can be made
finite. Following our discussion from sec.\ 5 we use the ordered limits
procedure, which, using (\ref{ghokl}), gives
\begin{eqnarray}
\lim_{\tilde{\epsilon}_1\to1} \left( \lim_{\tilde{\epsilon}_2\to1} 
 \langle\widehat{\widetilde{\Xi}}_{\tilde{\epsilon}_1}|
 \mathcal{Q} \,| \widehat{\widetilde{\Xi}}_{\tilde{\epsilon}_2}\rangle \right) 
 &=& \lim_{\tilde{\epsilon}_1\to1} \left[ 1 - 
 \frac{\tilde{\epsilon}_1}{1+(1-\tilde{\epsilon}_1)\tilde{\kappa}}
 \right]^2
 \frac{\det(1 - \widehat{\widetilde{S}}^2)}{\left[
 \Det(1 - \widetilde{\Sigma} \widetilde{\mathcal{V}})\right]^2}
 \nonumber \\
&=& \lim_{\tilde{\epsilon}_1\to1} 
 \left[ \frac{(\tilde{\kappa}+1)(1-\tilde{\epsilon}_1)}{1+
 (1-\tilde{\epsilon}_1)\tilde{\kappa}}\right]^2
 \frac{L^{\frac{11}{18}}}{\tilde h} + \ldots \;.
\end{eqnarray}
Therefore, if we assume that
\begin{equation}
1-\tilde{\epsilon}_1 = \tilde{s} L^{-\frac{11}{36}}
\end{equation}
for some constant $\tilde{s}$, we have
\begin{equation}
\lim_{\tilde{\epsilon}_1\to1} \left( \lim_{\tilde{\epsilon}_2\to1} 
\langle\widehat{\widetilde{\Xi}}_{\tilde{\epsilon}_1}|
\mathcal{Q} \, |\widehat{\widetilde{\Xi}}_{\tilde{\epsilon}_2}\rangle \right)
= (\widetilde{\kappa}+1)^2 \, \frac{\tilde{s}^2}{\tilde h}\;.\label{ghostlim}
\end{equation}
which defines a finite value for the kinetic term in the action.

The calculation for the cubic term in the action goes along similar
lines. One obtains here too that the ordered limits preserve the equation
of motion,
\begin{equation}
\lim_{\tilde{\epsilon}_1\to1} \left( \lim_{\tilde{\epsilon}_2\to1} 
\left(\lim_{\tilde{\epsilon}_3\to1}  
\langle\widehat{\widetilde{\Xi}}_{\tilde{\epsilon}_1}|
\widehat{\widetilde{\Xi}}_{\tilde{\epsilon}_2} *_g 
\widehat{\widetilde{\Xi}}_{\tilde{\epsilon}_3}\rangle \right) \right)
= -\lim_{\tilde{\epsilon}_1\to1} \left( \lim_{\tilde{\epsilon}_2\to1} 
\langle\widehat{\widetilde{\Xi}}_{\tilde{\epsilon}_1}|
 \mathcal{Q} \,| \widehat{\widetilde{\Xi}}_{\tilde{\epsilon}_2}\rangle \right)
\end{equation}

It is worth noting that, using the results of \cite{MM}, the ghost companion
of the dressed sliver can be easily shown to be (proportional to) a projector
of the  $bc$--twisted $*$--product \cite{GRSZ2}.

\subsection{Overall regularized action}

Now we are ready to draw the conclusion concerning the regularized
action. We collect the results (\ref{xi1xi2f},\ref{ghostlim})
and plug them into 
(\ref{actionsliver}). The action
of the regularized dressed sliver is 
\beq
-\frac{{\mathcal S}(\hat\Psi)}{V^{(D)}}= \frac 1{6g_0^2(2\pi)^D}\, \frac{(\tilde \k+1)^2{\tilde s}^2}
{(\k+1)^Ds^D}\,\frac {h^{\frac D2}}{{\tilde h}}\label{finaction}
\eeq
The value of the $rhs$ can now be tuned to the physical value of the 
D25--brane tension.
We stress that, apart from $g_0$, the parameters in the $rhs$ are not present
in the initial action, but arise from the regularization 
procedure\footnote{We remark that $\k$, $\tilde\k$, $h,\tilde h$
could be  reabsorbed in the free parameters $s$,$\tilde s$.}.  
More comments on this point can be found in section 8.

We would also like to point out that the regularized action 
(\ref{finaction})
is not the only possibility. We could, for instance,
connect in various ways the ghost and matter asymptotic expansions, to get
an overall  finite action.  We could perhaps
use also the limits $\k,\tilde \k\to -1$. At this stage we cannot
decide what  the best prescription is. Hopefully the study
of the spectrum will shed light on this problem.

\section{Other finite norm solutions.}
In this section we discuss a few further issues concerning dressed slivers,
without going into detailed calculations. 

\noindent $\bullet$ {\bf Multiply dressed slivers.}
The most obvious generalization of the dressed sliver definition 
(\ref{Xihat}) consists in
adding to $\hat S$ another operator $R'$ with the same structure as
$R$ and $\xi$ replaced by $\xi'$, with
\beq
\rho_1 \xi' =0,\quad\quad \rho_2 \xi' =\xi', \label{xi'}
\eeq
and
\beq
\xi'^T \frac 1{1-T^2}\xi' =1 ,\quad\quad
\xi'^T \frac {T}{1-T^2}\xi'= \kappa'
\label{noncond'}
\eeq
the components of $\xi'$ being real and $\k'$ a real number.
The matrix $\hat T$ will be replaced by
\beq
\hat T' = T +P+P',\quad\quad  P'=\frac 1{\kappa' +1}
\left(|\xi'\rangle\langle\xi'|+|C\xi'\rangle\langle C\xi'|
\right)\label{P'}
\eeq
The obvious question is whether this new state is a projector. In general
it is not, but if $\xi'$ satisfies the `orthogonality' conditions
\beq
\xi^T \frac 1{1-T^2}\xi' =0 ,\quad\quad
\xi^T \frac {T}{1-T^2}\xi'= 0
\label{orthog}
\eeq
then it is easy to repeat the proof of section 2 and conclude that
the squeezed state with structure matrix $\hat S'= S+R+R'$ is in fact a projector.
On the basis of section 3, one can see that the conditions
(\ref{orthog}) are easy to implement.

Again, the norm (and the action) of this new projector is ill--defined.
We can introduce deformation parameters $\e$ before $P$  and $\e'$ 
before $P'$, and repeat what we did in section 3, 4 and 5.
For instance, for $\e,\e'$ near 1, denoting by $\hat {S}_{\e,\e'}$ 
the relevant Neumann matrix,
\be
&&\det(1-\hat {S}_{\e,\e'}^2)= 
\det(1-T^2)(1-\e)^2(1-\e')^2\frac{16}{(\k+1)^2(\k'+1)^2}\label{S'2}\\
&&\Det(1-\hat\Sigma'_{\e,\e'}\V)=
\Det(1-\T\M)\frac 1{(\k+1)^2(\k'+1)^2}\label{S'V}
\ee
and so on. It is obvious that we can add to $\hat S$ as many perturbations
as we wish and still get projectors. For instance, if we add $R''$, 
specified by $\xi''$, with
the same properties as $\xi$, the only condition we have to impose is that
$\xi''$ be orthogonal to both $\xi$ and $\xi'$ in  the sense of 
eq.(\ref{orthog}).

\vskip .2cm
\noindent$\bullet$ {\bf Other projectors.}
Starting from the dressed sliver solutions it is rather easy to construct 
many others which are $*$--orthogonal to the dressed sliver, according
to the construction initiated in \cite{RSZ3} and fully implemented in
\cite{BMS3}.  
First we introduce a real vector $\zeta^\mu =
\{\zeta^\mu_{n}\}$ (notice the Lorentz index!), which is chosen to 
satisfy the conditions
\be
\rho_1 \zeta^\mu =0,\quad\quad \rho_2\zeta^\mu=\zeta^\mu, \quad\quad
\forall \mu
\label{zeta}
\ee
and
\be
\langle\zeta^\mu |\frac 1{1-T^2}|\zeta^\nu\rangle \eta_{\mu\nu} =1 ,
\quad\quad
\langle \zeta^\mu| \frac {T}{1-T^2}|\zeta^\nu\rangle\eta_{\mu\nu}= \l
\label{cond0}
\ee

Next we set
\be
\x = -(a^{\mu\dagger}  \zeta^\nu\,\eta_{\mu\nu})\, 
(a^{\mu\dagger} C \zeta^\nu\eta_{\mu\nu}),   \label{x}
\ee
introduce the Laguerre polynomials $L_n(\x/\l)$ and
define the states $|\hat\Lambda_n\rangle$ as follows
\be
|\hat\Lambda_n\rangle = (-\l)^n L_n\Big(\frac{\x}{\l}\Big) |\hat\Xi\rangle
\label{Lambdan}
\ee
where $\l$ is an arbitrary real constant, and $|\hat\Xi\rangle$ is the 
dressed sliver. 

If, in addition to the above conditions, $\zeta^\mu$ are `orthogonal'
to the dressing vector $\xi$,
 \be
\langle\zeta^\mu |\frac 1{1-T^2}|\xi\rangle =0 ,\quad\quad
\langle \zeta^\mu| \frac {T}{1-T^2}|\xi\rangle= 0, \quad\quad
{\rm for~~any~~}\mu, \label{cond0'}
\ee
it is not hard to generalize the proofs of \cite{RSZ3},\cite{BMS3}
and conclude that
\be
&& |\hat\Lambda_n\rangle * |\hat\Lambda_m\rangle = \delta_{n,m}
|\hat\Lambda_n\rangle \label{nstarm}\\
&& \langle \hat\Lambda_n |\hat\Lambda_m\rangle = \delta_{n,m}
 \langle \hat\Xi |\hat\Xi\rangle \label{nm}
\ee
As explained in section 3, the additional conditions (\ref{cond0'})
are easy to comply with.

\vskip .2cm
\noindent$\bullet$ {\bf Lump solutions.}
In VSFT lump solutions of any dimension have been found, \cite{RSZ2, RSZ3}.
They are candidates to represent lower dimensional branes. 
By definition, they are not translational invariant in a subset of 
directions (the transverse ones). In order to find such solutions
we cannot drop anymore the momentum dependence in the transverse
directions. We therefore proceeds as follows. 
First we split the Lorentz indices $\mu,\nu$ into parallel ones,
$\bar \mu,\bar \nu$, running from $0$ to $25-k$, and transverse ones,
$\a,\b$ which run from $26-k$ to 25. Next we introduce the zero modes
by defining
\beq
a_0^{(r)\alpha} = \frac 12 \sqrt b \hat p^{(r)\alpha}
- i\frac {1}{\sqrt b} \hat x^{(r)\alpha},
\quad\quad
a_0^{(r)\alpha\dagger} = \frac 12 \sqrt b \hat p^{(r)\alpha} +
i\frac {1}{\sqrt b}\hat x^{(r)\alpha}, \label{osc}
\eeq
where $\hat p^{(r)\alpha}, \hat x^{(r)\alpha}$ are the momentum
and position operator of the $r$--th string. The parameter
$b$ is as in ref.\cite{RSZ2}. 	
The Dirac brackets for all the oscillators including the zero modes
are, in the transverse directions,
\beq
\big[a_M^{(r)\alpha},a_N^{(s)\beta\dagger}\big]=
\eta^{\alpha\beta}\delta^{rs}\delta_{MN},\quad\quad\quad N,M\geq 0
\label{a0a0}
\eeq
where the index $N$ denotes the couple $(0,n)$.
Now we ontroduce $|\Omega_{b}\rangle$, the oscillator vacuum
(\,$a_N^\alpha|\Omega_{b}\rangle=0$, for $N\geq 0$\,).
The relation
between the momentum basis and the oscillator basis is defined by
\be
|\{p^\alpha\}\rangle_{123}
=\left(\frac b{2\pi}\right)^\frac k4 e^{\sum_{r=1}^3
\left(- \frac b4 p^{(r)}_\alpha \eta^{\alpha\beta}p^{(r)}_\beta+
\sqrt b  a_0^{(r)\alpha\dagger}p^{(r)}_\a
- \frac 12 a_0^{(r)\alpha\dagger}\eta_{\alpha\beta}a_0^{(r)\beta\dagger}
\right)}|\Omega_b\rangle\0
\ee
Inserting this into (\ref{V3}) and integrating with respect to the
transverse momenta one finally gets the following three strings
vertex \cite{GJ1,RSZ2}
\beq
|V_3 \rangle' = |V_{3,\perp}\rangle ' \,\otimes\,|V_{3,\|}\rangle
\label{split'}
\eeq
$|V_{3,\|}\rangle$ is the one used before this subsection
(with zero momenta), while
\beq
|V_{3,\perp}\rangle'= K_2\, e^{-E'}| \Omega_b\rangle\label{V3'}
\eeq
where $K_2$ is a suitable constant and
\beq
E'= \frac 12 \sum_{r,s=1}^3 \sum_{M,N\geq 0} a_M^{(r)\a\dagger}
{V'}_{MN}^{rs} a_N^{(s)\b\dagger}\eta_{\a\b}\label{E'}
\eeq

The vertex coefficients ${V'}_{MN}^{rs}$ to be used in the transverse 
directions have parallel properties to the vertex $V_{mn}^{rs}$.
When multiplied by the twist matrix they give rise to matrices
$X',X'_+,X'_-$ which happen to obey the same equations collected in 
Appendix A
for the matrices $X,X_+,X_-$. Since all the calculations we have done
throughout the paper depend uniquely on such properties, we can repeat
everything almost {\it verbatim}. So, there will be a matrix $T'$ given
by a formula (\ref{sliver}), with a normalization (\ref{norm}) and a
$bpz$ norm (\ref{ener}), where all the entries are primed.
Next we introduce the dressed sliver exactly as in section 2.
To this end first we define the infinite vector $\xi'= \{\xi'_N\}$
satisfying the condition
\beq
\rho'_1 \xi' =0,\quad\quad \rho'_2 \xi' =\xi', \label{xi''}
\eeq
and 
\beq
\xi'^T \frac 1{1-{T'}^2}\xi'=1 ,\quad\quad
\xi'^T \frac {T'}{1-{T'}^2}\xi'= \kappa
\label{noncond''}
\eeq
where $\kappa$ is the same number as in (\ref{noncond}). The
transverse dressed sliver is defined by
\beq
|\hat\Xi_\perp\rangle = \hat{\cal N'} e^{-\frac 12 a^\dagger \hat S' a^\dagger}| \Omega_b\rangle\,
\label{Xihat'}
\eeq
where
\beq
\hat S' = S' +R',\quad\quad R'_{MN}= \frac 1{\kappa +1}
\left(\xi'_M(-1)^N\xi'_N
+\xi'_N(-1)^M\xi'_M\right)\label{Shat'}
\eeq
and so on. The proofs of section 3 can be repeated, given the diagonal 
structure of Neumann matrices with zero modes \cite{Feng}.
Once again we introduce a deformation parameter $\e$ (the same as in
section 4!) and repeat the derivations of section 5 (where D,  
for the transverse directions, equals $k-1$).

One of the most remarkable results of VSFT is the reproduction
of the ratio of tensions for brane of different dimensions.
It is important to verify that our regularization procedure
does not alter this ratio. 

It is easy to show that in the present case the ratio of tensions
for brane of adjacent dimensions can be written as follows
\beq
\frac{\T_{24-k}}{2\pi \T_{25-k}} = \frac 3{\sqrt{2\pi b^3}} \,\left(
{V_{00} +\frac b2}\right)^2\,\frac{(\det(1-X')^{3/4}\det(1+3X')^{1/4}}
{(\det(1-X)^{3/4}\det(1+3X)^{1/4}}\cdot f(\e,\k)\label{ratio}
\eeq
The factor $f(\k,\e)$ is due to dressing. However it is elementary
to prove that this factor is actually 1. What remains is the same
as in \cite{RSZ2}. It was proven  numerically \cite{RSZ2} and 
analytically \cite{Oku3} that the ratio at the $rhs$ of (\ref{ratio}) 
is exactly 1, thus reproducing the expected ratio.

It goes without saying that one can easily introduce a constant background
$B$ field in the transverse directions, along the lines of \cite{BMS1,
BMS2}.

\section{Some comments and conclusions}

There has been in the recent past some debate concerning the
possibility to define finite
action solutions in VSFT, \cite{GRSZ1,Sch1,Oka2}.  
Here we would like to clarify the relation of our treatment with these
previous works. First of all our 
action does not require any overall infinite normalization, while
the ones proposed by \cite{RSZ1, HKw} are  singularly 
normalized.

Let us begin with the VSFT action proposed by Hata and Kawano \cite{HKw}
\beq\label{Shk1}
S_{HK}[\psi]=-K\left(\frac12\bra\psi\Q\ket\psi+
\frac13\bra\psi\psi*\psi\rangle\right)
\eeq
where, from both conformal and operatorial considerations, 
$\Q$ is given by \cite{GRSZ1, HKw, Oku2}
\be
\Q=\frac{1}{2i}\left(c(i)-c(-i)\right)\0
\ee
In \cite{GRSZ1} Gaiotto {\it et al.} argued that a $\Q$ of this kind should be 
obtained by a (singular) field redefinition, so that the OSFT action 
expanded on the tachyon vacuum would take the form
\beq\label{Srsz}
S_{GRSZ}[\tilde\psi]=-\frac{1}{g_0^2}\left(\frac{1}
{2\e}\bra{\tilde\psi}\Q\ket{\tilde\psi}+\frac13\bra{\tilde\psi}
\tilde\psi*\tilde\psi\rangle\right)
\eeq 
where an $\e\rightarrow0$ limit is understood. Another  numerical 
field redefinition $\tilde\psi=\left(g_0^2K\right)^{1/3}\psi$ brings  
$S_{GRSZ}$ to $S_{HK}$, provided we identify
\beq
K=\frac{1}{g_0^2\e^3}
\eeq
so that
\beq\label{Shk2} 
S_{HK}[\psi]=-\frac{1}{g_0^2\e^3}\left(\frac12\bra\psi\Q\ket\psi+
\frac13\bra\psi\psi*\psi\rangle\right)
\eeq
In the present work we have instead considered a finitely normalized 
action
\beq\label{Sbmp}
S[\hat\psi]=-\frac{1}{g_0^2}\left(\frac{1}{2}\bra{\hat\psi}\Q\ket{\hat\psi}+
\frac13\bra{\hat\psi}\hat\psi*\hat\psi\rangle\right)
\eeq 
This  action can be indeed brought to the $S_{HK}=S_{GRSZ}$ form by the operatorial 
field redefinition \cite{Oka2}
\beq\label{redef}
\psi=e^{-\frac14\ln\e(K_2-4)}\hat\psi
\eeq
In fact the family of operators $K_n=L_n-(-1)^nL_{-n}$ leaves the cubic 
term invariant while it acts non trivially on the kinetic term as
\beq
[K_{2n},\Q]=-4n(-1)^n\Q\label{KQ}
\eeq
So, if a  classical solution exists that can render our VSFT action finite  
(that is, such that we can find a regularization scheme in which the $bpz$--norm can 
be tuned to a finite value), then something similar should be done with the 
singularly normalized actions (\ref{Srsz}, \ref{Shk2}). This is in fact 
what Okawa found in \cite{Oka2} using the regularized (twisted) butterfly 
state \cite{GRSZ1, GRSZ2} and representing it in the non--twisted CFT.  
In this way he showed that one can even reabsorb the open string 
coupling constant into a field redefinition of the kind (\ref{redef}), so 
that such a coupling constant is not a free parameter.

This fact seems to mark a difference between OSFT and VSFT. In the former
one cannot implement a redefinition like (\ref{redef}) keeping the
ratio of the kinetic and interaction terms constant.
Therefore in VSFT there seems to be no free parameter in the action.
It is not clear to us if such a feature of VSFT is really a problem:
in fact we have seen that free parameters appear effectively in the 
process of regularizing the classical solution, and one should  note 
that VSFT gives the right ratios between energy densities \cite{Oku3, Oka}. 
So, if a finite energy solution depending on such effective parameters 
is defined in such a way that they are constrained to yield the correct 
physical value for its tension,
the tensions of all the other branes are correctly predicted. 

Finally, our results in this paper may shed some light on a conclusion
drawn in \cite{Oka2} concerning the validity of the EOM inside the action.
From the formulas of Appendix C one can see that, if one    
insists in defining the dressed sliver as the strong 
limit of $\hat\Psi_{\e,\tilde\e}=\hat\Xi_\e\otimes 
\widehat{\widetilde\Xi}_{\tilde\e}$ for $\e,\tilde\e\to 1$, it follows that
\beq
\lim_{\e\to 1,\tilde\e\to 1}\bra{\hat\Psi_{\e,\tilde\e}}\Q|
\hat\Psi_{\e,\tilde\e}\rangle=-
\left(\frac32\right)^{D-2}\lim_{\e\to 1,\tilde\e\to 1}
\bra{{\hat\Psi}_{\e,\tilde\e}}\hat\Psi_{\e,\tilde\e}*
\hat\Psi_{\e,\tilde\e}\rangle\label{misma}
\eeq
An analogous problem is shown to exist in \cite{Oka2}. It was supposed 
there
that, in order to satisfy the EOM one should take into account
subleading terms in $\Q$, hence loosing the very appealing matter--ghost 
factorization of VSFT. Here we have instead attributed the
mismatch (\ref{misma}) to subtleties of the limiting procedure.
\subsection{Conclusions}
In this paper we have shown that it is possible to find solutions of 
VSFT with finite $bpz$--norm and action. This result was achieved by first
introducing a new kind of solutions of VSFT, which we called dressed
slivers. The latter is a deformation of the well--known
sliver solution by the addition of a rank one projector to the
Neumann matrix. The dressed sliver, introduced in this naive way, has still
an ill-defined norm (and action), but one can naturally
introduce a regularization parameter $\e$ (which interpolates between 
the sliver and the dressed sliver), and tune it to the level truncation
parameter $L$. This leads to a finite $bpz$--norm and action.
We have seen that this regularization is far from unique.
However we have also seen that most regularization prescriptions are
incompatible with the EOM. Requiring the validity of the EOM we have 
been able to define a unique prescription. We leave to \cite{BMP2} the problem
of testing this prescription on the open string states of the dressed sliver.
\acknowledgments
We would like to thank M.Schnabl and D.Mamone for useful discussions
and suggestions.
C.M. would like to thank L.F.Alday and M.Cirafici for discussions.
P.P. would like to thank SISSA-ISAS (Trieste) and Dipartimento
di Fisica Teorica (Universit\`{a} degli Studi di Torino) for their kind 
hospitality and stimulating atmosphere during this research.
This research was supported by the Italian MIUR
under the program ``Teoria dei Campi, Superstringhe e Gravit\`a''.

\section*{Appendix}
\appendix
\section{A collection of well--known formulae}
In this Appendix we collect some useful results and formulas involving the
matrices of the three strings vertex coefficients.

To start with, we recall that
\begin{itemize}
\item (i) $V_{nm}^{rs}$ are symmetric under simultaneous exchange of
the two couples of indices;
\item (ii) they are endowed with the property of cyclicity in the
$r,s$ indices, i.e. $V^{rs}= V^{r+1,s+1}$, where $r,s=4$ is
identified with $r,s=1$.
\end{itemize}

Next, using the twist matrix $C$  ($C_{mn}= (-1)^m \delta_{mn}$), we define
\beq
X^{rs} \equiv C V^{rs}, \quad\quad r,s=1,2,\label{EX}
\eeq
These matrices are often rewritten in the following way $X^{11}=X,\, X^{12}=X_+,\, 
X^{21}=X_-$. They commute with one another
\beq
[X^{rs}, X^{r's'}] =0, \label{commute}
\eeq
moreover
\beq
CV^{rs}= V^{sr}C ,\quad\quad CX^{rs}= X^{sr}C
\eeq
Next we quote some useful identities:
\be
&&X+ X_++ X_- = 1\0\\
&& X_+X_- = (X)^2-X\0\\
&& (X_+)^2+ (X_-)^2= 1- (X)^2\0\\
&& (X_+)^3+ (X_-)^3 = 2 (X)^3 -  3(X)^2 +1
\label{Xpower}
\ee
and
\beq
\frac{1-TX}{1-X}=\frac 1{1-T},\quad\quad \frac{X}{1-X}=\frac T{(1-T)^2}
\label{TXid}
\eeq
Using these one can show, for instance, that
\be
&& \K^{-1} = \frac 1{(1+T)(1-X)} \left(\matrix {1-TX & TX_+\cr TX_- &1-TX\cr}
\right)\0\\
&&\M\K^{-1} =\frac 1{(1+T)(1-X)} \left(\matrix {(1-TX)X  &X_+ \cr
X_-&(1-TX)X\cr}\right) \label{id1}
\ee

Another ingredient we need is given by the Fock space projectors
\be
\rho_1 \!&=&\! \frac 1{(1 +T)(1-X)} \left[ X_+ (1-TX)
+T (X_-)^2\right]\label{rho1}\\
\rho_2 \!&=&\! \frac 1{(1 +T)(1-X)} \left[ X_- (1-TX)
+T (X_+)^2\right]\label{rho2}
\ee
They satisfy
\beq
\rho_1^2 = \rho_1,\quad\quad \rho_2^2 = \rho_2, \quad\quad 
\rho_1+\rho_2 = 1, \quad\quad \rho_1\rho_2 = 0\label{proj12}
\eeq
i.e. they project onto orthogonal subspaces. Moreover,  
\beq
\rho_1^T=\rho_1 = C\rho_2 C,\quad\quad
\rho_2^T=\rho_2 = C\rho_1 C.\label{rhorels}
\eeq
where $^T$ represents matrix transposition. As was shown in
\cite{RSZ3,Moeller}, $\rho_1,\rho_2$ projects out half the string modes.
Using these projectors one can prove that
\be
(X_+,X_-)\,\K^{-1} = (\rho_1,\rho_2),\quad\quad
\M \K^{-1} \T \left(\matrix {X_-\cr X_+\cr}\right) =
\left(\matrix {TX\rho_2 +TX_+\rho_1\cr TX_-\rho_2+TX\rho_1\cr}\right)
\label{id2}
\ee
which are used throughout the paper.

The following relations are often useful
\beq
\rho_1X_+ +\rho_2X_- = 1-XT,\quad\quad \rho_1X_- +\rho_2X_+ =X(T-1)
\label{Xrho}
\eeq

In the ghost sector matrices $\widetilde{X}$, $\widetilde{X}_\pm$
satisfy all relations above. In particular one can define
the half--string projectors $\tilde\rho_1,\tilde\rho_2$ as in (\ref{rho1},\ref{rho2}) and
verify that they satisfy the same relations as the matter projectors. 
 For the manipulations with zero-modes
it is useful here to define the vectors
\begin{equation}
(\widetilde{\mathbf{v}}_0)_n = \widetilde{V}_{n0}^{r,r} \;, \qquad
(\widetilde{\mathbf{v}}_\pm)_n = \widetilde{V}_{n0}^{r,r\pm1}
\end{equation}
which satisfy
\begin{equation}
\widetilde{\mathbf{v}}_0 = (1-\widetilde{X}) \mathbf{f} \;, \qquad
\widetilde{\mathbf{v}}_\pm = - \widetilde{X}_{\mp} \mathbf{f}
\end{equation}
where $\mathbf{f}=\{f_n\}$ is given by
\begin{equation}
f_n = \cos\left(\frac{n\pi}{2}\right) \;.
\end{equation}
This vector $\mathbf{f}$ appears in the expression for
the kinetic operator $\mathcal{Q}$:
\begin{equation}
\mathcal{Q} = c_0 + \sum_{n=1}^{\infty} f_n 
\left( c_n + (-1)^n c_n^\dagger \right) \;.
\end{equation}

\section{Evaluation of determinants}
This section is devoted to the evaluation of determinants which appear 
in calculations involving dressed slivers. Here we deal only with the
matter determinants, but the same results hold for the corresponding
ghost determinants.

\subsection{$\Det(1-\hat\T_\e\M)$}
First of all we consider
\beq
(1-\hat\T_\e\M)^{-1}\CP=\K^{-1}(1-\e \CP\M\K^{-1})^{-1}\CP
\eeq
This matrix can be exactly computed from
\beq
\CP\M\K^{-1}\CP= \left(\matrix{\k&\rho_1-\kappa\rho_2\cr
\rho_2-\kappa\rho_1&\k\cr}\right)\CP
\eeq
We have in fact
\beq
(1-\e \CP\M\K^{-1})^{-1}\CP=\sum_{n=0}^{\infty}\left(\e \CP\M\K^{-1}\right)^n\CP
\eeq
Using the properties of the $\rho$ projectors, defined in the previous 
appendix, we can easily show that
\beq
\left(\e \CP\M\K^{-1}\right)^n\CP=\frac{\e^n}{(\k+1)^n}    
\left(\matrix{  
A(n)
& 
B(n)(\rho_1-\kappa\rho_2)       
 \cr 
B(n)(\rho_2-\kappa\rho_1)   
&
A(n)
 \cr}\right)\CP
\eeq
where
\be
A(n)&=&\sum_{l=0}^{\left[\frac n2\right]} (-1)^l k^{n-l}\left(\matrix {n \cr 2l\cr}\right)\\
B(n)&=&\sum_{p=0}^{\left[\frac{n-1}{2}\right]} (-1)^p k^{n-p-1}\left(\matrix {n \cr 2p+1\cr}\right)
\ee
Now we exchange the order of summations
\be
\sum_{n=0}^{\infty}\sum_{l=0}^{\left[\frac n2\right]}&=&\sum_{l=0}^{\infty}\sum_{n=2l}^{\infty}\0\\
\sum_{n=0}^{\infty}\sum_{p=0}^{\left[\frac{n-1}{2}\right]}&=&\sum_{p=0}^{\infty}\sum_{n=2p+1}^{\infty}\0
\ee
and use the resummation formula
\beq
\sum_{n=l}^{\infty}\left(\matrix {n \cr l\cr}\right)\frac{p^{n-l}}{q^n}=\frac{q}{(q-p)^{l+1}}
\eeq
With standard algebraic manipulations, we get
\beq
(1-\e\CP\M\K^{-1})^{-1}\CP=\frac{\k+1}{(\k-\e\k+1)^2+\e^2\k}\left(\matrix{  
\k-\e\k+1
& 
\e(\rho_1-\kappa\rho_2)       
 \cr 
\e(\rho_2-\kappa\rho_1)   
&
\k-\e\k+1
 \cr}\right)\CP
\eeq
In order to compute $\Tr\ln(1-\e\CP\M\K^{-1})$ we first consider
\be\label{dde}
&&\frac{d}{d\e}\Tr\ln(1-\e\CP\M\K^{-1})=
-\Tr\left[(1-\e\CP\M\K^{-1})^{-1}\CP\M\K^{-1}\right]\0\\
\0\\
&&=-\Tr\left[\frac{\k+1}{(\k-\e\k+1)^2+\e^2\k}
\left(\matrix{  \k-\e\k+1 & \e(\rho_1-\kappa\rho_2)
 \cr \e(\rho_2-\kappa\rho_1)   
& \k-\e\k+1 \cr}\right)\CP\M\K^{-1}\right]\0\\
\0\\
&&=-\frac{\k+1}{(\k-\e\k+1)^2+\e^2\k}
\tr\left[2(\k-\e\k+1)P\frac{T}{1-T^2}-\e P\frac{T}{1-T^2}-
\e\k P\frac{1}{1-T^2}\right]\0\\
\0\\
&&=-\frac{2(\k+1)}{(\k-\e\k+1)^2+\e^2\k}\left(2(\k-\e\k+1)
\frac{\k}{\k+1}-\e\frac{\k}{\k+1}-\e\k\frac{1}{\k+1}\right)\0\\
\0\\
&&=-4\frac{\k(\k+1)(1-\e)}{(\k-\e\k+1)^2+\e^2\k}
\ee
Hence we get
\beq
\Tr\ln(1-\e\CP\M\K^{-1})=-4\int_0^\e d\e' \, 
\frac{\k(\k+1)(1-\e')}{(\k-\e'\k+1)^2+\e'^2\k}=
2\ln\frac{1+(1-\e)^2\k}{\k+1}
\eeq
Collecting all the contributions we  finally obtain
\beq
\Det(1-\hat\T_\e\M)=\left(\frac{1+(1-\e)^2\k}{\k+1}\right)^2
\Det(1-\T\M)
\eeq

\subsection{$\Det(1-\hat\T_{\e_1\e_2}\M)$}
To compute this determinant we use the same strategy as 
before, that is we first compute 
\beq
(1-\CP_{\e_1\e_2}\M\K^{-1})^{-1}\CP_{\e_1\e_2}=\left(\matrix{  
A & B(\rho_1-\kappa\rho_2) \cr D(\rho_2-\kappa\rho_1)  
 & D \cr}\right)\CP_{\e_1\e_2}
\eeq
where $A,B,C,D$ are to be determined. Moreover we have defined
\be
\CP_{\e_1\e_2}&=&\left(\matrix{\e_1 & 0 \cr 0& \e_2}\right)P\\
\hat\T_{\e_1\e_2}&=&\left(\matrix{\hat T_{\e_1} & 0 \cr 0& \hat T_{\e_2}}\right)
\ee
The constant $A,B,C,D$ can be easily determined by imposing 
$(1-\CP_{\e_1\e_2}\M\K^{-1})^{-1}\CP_{\e_1\e_2}$ to  give
 back $\CP_{\e_1\e_2}$ when multiplied on the left by 
$(1-\CP_{\e_1\e_2}\M\K^{-1})$. The procedure is 
straightforward and gives the result
\beq\label{1-P12mk}
(1-\CP_{\e_1\e_2}\M\K^{-1})^{-1}\CP_{\e_1\e_2}=
\frac{1}{1+(1-\e_1)(1-\e_2)\k}\left(\matrix{  
\k+1-\e_2\k & \e_1(\rho_1-\kappa\rho_2) \cr \e_2(\rho_2-\kappa\rho_1)  
 &   \k+1-\e_1\k \cr}\right)\CP_{\e_1\e_2}
\eeq
Now we come to the computation of the determinant
\beq\label{1-t12m}
\Det(1-\hat\T_{\e_1\e_2}\M)=\Det(1-\T\M)\,
\exp\left(\Tr\ln(1-\CP_{\e_1\e_2}\M\K^{-1})\right)
\eeq
To compute the exponent of the second factor in  the $rhs$ we 
use the same strategy as before, namely we consider
\be
&&\frac{d}{dx}\Tr\ln(1-x\CP_{\e_1\e_2}\M\K^{-1})=
-\Tr\left[(1-x\CP_{\e_1\e_2}\M\K^{-1})^{-1}\CP_{\e_1\e_2}\M\K^{-1}\right]\0\\
\0\\&&=-\Tr\left[\frac 1x(1-\CP_{x\e_1,x\e_2}\M\K^{-1})^{-1}
\CP_{x\e_1,x\e_2}\M\K^{-1}\right]\\ \0\\
&&=-\Tr\left[\frac 1x \frac{1}{1+(1-\e_1)(1-\e_2)\k}\left(\matrix{ 
 \k+1-x\e_2\k & x\e_1(\rho_1-\kappa\rho_2) \cr x\e_2(\rho_2-
\kappa\rho_1)   &   \k+1-x\e_1\k \cr}\right)\CP_{x\e_1,x\e_2}\M\K^{-1}
\right]\0\\ \0\\
&&=-2\frac{(\e_1+\e_2)\k-2x\e_1\e_2\k}{1+(1-x\e_1)(1-x\e_2)\k}
\ee
where the same manipulations as in (\ref{dde}) have been used.
Then we perform the simple integration
\beq
\Tr\ln(1-\CP_{\e_1\e_2}\M\K^{-1})=-2\int_0^1dx 
\frac{(\e_1+\e_2)\k-2x\e_1\e_2\k}{1+(1-x\e_1)(1-x\e_2)\k}=
2\ln\left(\frac{1+(1-\e_1)(1-\e_2)\k}{\k+1}\right)
\eeq
Therefore we have obtained
\beq
\Det(1-\hat\T_{\e_1\e_2}\M)=\left(\frac{1+(1-\e_1)(1-\e_2)\k}{\k+1}
\right)^2\Det(1-\T\M)\label{1-T12M}
\eeq

\subsection{$\det(1-\hat T_\e^2)$}
We have
\beq\label{1-t2}
\det(1-\hat T_\e^2)=\det(1-\hat T_\e)\det(1+\hat T_\e)
\eeq
We compute the two factors separately
\beq
\det(1-\hat T_\e)=\det\left(1-\frac{\e}{1-T}P\right)\,\det(1-T)
\eeq
For the first factor in the $rhs$ we have
\be\label{1-t}
\det(1-\frac{\e}{1-T}P)&=&\exp\left(\tr\ln(1-\frac{\e}{1-T}P)\right)\0\\
&&=\exp\left(-\sum_{n=1}^{\infty}\frac 1n \tr(\frac{\e}{1-T}P)^n\right)
=\exp\left(-2\sum_{n=1}^{\infty}\frac {\e^n}{n(\k+1)^n}\bra\xi\frac{1}
{1-T}\ket\xi^n\right)\0\\
&&=\exp\left(-2\sum_{n=1}^{\infty}\frac {\e^n}{n(\k+1)^n}(\k+1)^n\right)
=\exp\left(2\ln(1-\e)\right)\0\\
&&=(1-\e)^2
\ee
So we have
\beq\label{1-T}
\det(1-\hat T_\e)=(1-\e)^2\,\det(1-T)
\eeq
Now let's turn to the second factor in (\ref{1-t2})
\beq
\det(1+\hat T_\e)=\det\left(1+\frac{\e}{1+T}P\right)\,\det(1+T)
\eeq
Computing as in (\ref{1-t}) we obtain
\beq\label{1+t}
\det\left(1+\frac{\e}{1+T}P\right)=\left(\frac{\k+1-\e(\k-1)}{\k+1}
\right)^2
\eeq
giving the result
\beq\label{1+T}
\det(1+\hat T_\e)=\left(\frac{\k+1-\e(\k-1)}{\k+1}\right)^2\,\det(1+T)
\eeq
Collecting the two results (\ref{1-t},\ref{1+t}) we get
\beq\label{1-T2}
\det(1-\hat T_\e^2)=(1-\e)^2\left(\frac{\k+1-\e(\k-1)}{\k+1}\right)^2\,
\det(1-T^2)
\eeq

\subsection{$\det(1-\hat T_{\e_1}\hat T_{\e_2})$}
First of all we decompose
\be
&&1-\hat T_{\e_1}\hat T_{\e_2}=(1-\hat T_{\e_1})(1+\hat T_{\e_2})+\hat T_{\e_1}-\hat T_{\e_2}\0\\
&&=(1-\hat T_{\e_1})(1+(\e_1-\e_2)(1-\hat T_{\e_1})^{-1}P(1+\hat T_{\e_2})^{-1})(1+\hat T_{\e_2})
\ee
So we have
\beq\label{dec}
\det(1-\hat T_{\e_1}\hat T_{\e_2})=\det(1-\hat T_{\e_1})\det(1+\hat T_{\e_2})\det(1+(\e_1-\e_2)
(1-\hat T_{\e_1})^{-1}P(1+\hat T_{\e_2})^{-1})
\eeq
We need to compute the third factor in $rhs$
\be\label{e1e2}
&&\det\left(1+(\e_1-\e_2)(1-\hat T_{\e_1})^{-1}P(1+\hat T_{\e_2})^{-1}\right)=\\
&&=\exp\left(\tr\ln(1+(\e_1-\e_2)(1-\hat T_{\e_1})^{-1}P(1+\hat T_{\e_2})^{-1})\right)\0\\
&&=\exp\left(-2\sum_{n=1}^{\infty}\frac{(-1)^n}{n}\left(\frac{\e_1-\e_2}{\k+1}\right)^n \bra\xi
(1+\hat T_{\e_2})^{-1}(1-\hat T_{\e_1})^{-1}\ket\xi^n\right)\0
\ee
where the factor 2 is to take into account the (equal) contributions of 
$\xi$ and $C\xi$ which constitute $P$, so from  now on only the 
contribution of $\xi$ is needed to be considered.
Then we decompose $(1+\hat T_{\e_2})^{-1}$ and 
$(1-\hat T_{\e_1})^{-1}$ as
\be
&&(1+\hat T_{\e_2})^{-1}=\left((1+\e_2P\frac{1}{1+T})(1+T)\right)^{-1}=
\frac{1}{1+T}\sum_{m=0}^{\infty}\left(\frac{-\e_2}{\k+1}\right)^m\left(\ket\xi\bra\xi\frac{1}{1+T}\right)^m\0\\
&&(1-\hat T_{\e_1})^{-1}=\left((1-T)(1+\frac{1}{1-T}\e_1P)\right)^{-1}=
\sum_{p=0}^{\infty}\left(\frac{\e_1}{\k+1}\right)^p\left(\frac{1}{1-T}(\ket\xi\bra\xi\right)^p\frac{1}{1-T}\0
\ee
So we get
\be
&&\bra\xi(1+\hat T_{\e_2})^{-1}(1-\hat T_{\e_1})^{-1}\ket\xi^n\0\\
&&=\left(\sum_{m=0}^{\infty}\left(\frac{-\e_2}{\k+1}\right)^m\bra\xi\frac{1}{1+T}\ket\xi^m\bra\xi\frac{1}{1-T^2}\ket\xi\sum_{p=0}^{\infty}\left(\frac{\e_1}{\k+1}\right)^p\bra\xi\frac{1}{1-T}\ket\xi^p\right)^n\0\\
&&=\left(\frac{\k+1}{(\k+1-\e_2(\k-1))(1-\e_1)}\right)^n
\ee
plugging this in (\ref{e1e2}), we get
\be
&&\det\left(1+(\e_1-\e_2)(1-\hat T_{\e_1})^{-1}P(1+\hat T_{\e_2})^{-1}\right)\\
&&=\exp\left(-2\sum_{n=1}^{\infty} \frac1n \left(\frac{\e_2-\e_1}{(1-\e_1)(\k+1-\e_2(\k-1)}\right)^n\right)\0\\
&&=\left(1-\frac{\e_2-\e_1}{(1-\e_1)(\k+1-\e_2(\k-1)}\right)^2
\ee
From (\ref{dec}), using (\ref{1-T}, \ref{1+T}), we finally get

\be\label{1-t1t2}
&&\det(1-\hat T_{\e_1}\hat T_{\e_2})=\det(1-T^2)\left(1-(\e_1+\e_2)\frac{\k}{\k+1}+\e_1\e_2\frac{\k-1}{\k+1}\right)^2\\
&&=\det(1-T^2)\left[\frac{\e_1\e_2}{\k+1}\left(1-\frac{1}{\e_1\star\e_2}\right)\right]^2\0\\
&&=\det(1-\hat T_{\e_1}^2)^{\frac12}\det(1-\hat T_{\e_2}^2)^{\frac12}\left(1-\frac{\e_2-\e_1}
{(1-\e_1)(\k+1-\e_2(\k-1)}\right)\left(1-\frac{\e_1-\e_2}{(1-\e_2)(\k+1-\e_1(\k-1)}\right)\0
\ee
Note that the last two factors in $rhs$ of the last line  approach 1 as 
$\e_1\rightarrow\e_2$.

\section{Limit prescriptions}

\subsection{Double limit}

In this appendix we analyse various limits of the quantity
\beq
\langle \hat \Xi_{\e_1} |\hat \Xi_{\e_2} \rangle = 
\frac {\hat \N_{\e_1}
\hat \N_{\e_2}}{\det(1-\hat S_{\e_1}\hat S_{\e_2})}\langle 0|0\rangle 
\label{X1X2'}
\eeq
when $\e_1,\e_2\to 1$. We recall that $0\leq \e_1,\e_2\leq 1$.

Since
$\det(1-\hat S_{\epsilon_1} \hat S_{\epsilon_2})=\det(1-\hat S_{\epsilon_2} \hat S_{\epsilon_1})$
and
\beq
\det(1-\hat S_{\epsilon_1} \hat S_{\epsilon_2}) = 
\left( \det(1-\hat T_{\epsilon_1})
\det(1+\hat T_{\epsilon_2})
\det\left(1+({\epsilon_1}-{\epsilon_2})P
\frac 1{(1-\hat T_{\epsilon_1})(1+\hat T_{\epsilon_2})}\right)
\right)^\frac 12\0
\eeq
it is convenient to symmetrize the result. One gets
\be
&&\det(1-\hat S_{\epsilon_1} \hat S_{\epsilon_2})=
\left(\det(1-\hat T_{\epsilon_1}^2)
\det(1-\hat T_{\epsilon_2}^2)\right)^{\frac 12}\label{S1S2}\\
&&~~~\cdot \left(\det\left(1+({\epsilon_1}-
{\epsilon_2})P\frac 1{(1-\hat T_{\epsilon_1})(1+\hat T_{\epsilon_2})}\right)
\det\left(1+({\epsilon_2}-{\epsilon_1})P
\frac 1{(1-\hat T_{\epsilon_2})
(1+\hat T_{\epsilon_1})}\right)\right)^\frac 12\0
\ee
Using the results of Appendix B.4  this can be rewritten as
\be\label{S1S2'}
&&\det(1-\hat S_{\epsilon_1} \hat S_{\epsilon_2}) 
=\det(1-\hat T_{\e_1}^2)^{\frac12}
\det(1-\hat T_{\e_2}^2)^{\frac12}\0\\
&&~~~~~~~~~~\cdot\left(1-\frac{\e_2-\e_1}{(1-\e_1)
(\k+1-\e_2(\k-1)}\right)\left(1-\frac{\e_1-\e_2}
{(1-\e_2)(\k+1-\e_1(\k-1)}\right)\0
\ee
Therefore, collecting the previous results,
\be
&&\frac 1{\langle 0|0\rangle}\,\langle \hat \Xi_{\e_1} |\hat \Xi_{\e_2} \rangle = 
\left(\frac {\hat \N_{\e_1}}
{\sqrt{\det(1-\hat S_{\e_1}^2)}}\right)^{\frac D2}
\left(\frac {\hat \N_{\e_2}}
{\sqrt{\det(1-\hat S_{\e_2}^2)}}\right)^{\frac D2}\0\\
&&~~~~~~~~~\cdot\left(\frac {(1-\e_1)(1-\e_2)(\k+1-\e_1(\k-1))
(\k+1-\e_2(k-1))}
{(\k+1-(\e_1+\e_2)\k+\e_1\e_2(\k-1))^2}\right)^{\frac D2}
\label{xi1xi2}
\ee
When $\e_1$ and $\e_2$ are in the vicinity of 1, this simplifies 
as follows
\beq\label{xi1xi2'}
\left(\frac {\det(1-\Sigma\V)}{\sqrt{\det(1-S^2)}}\right)^D
\left(\frac 1{4(\k+1)^2}\right)^{\frac D2}
\left(\frac{4}{(\k(1-\e_1)(1-\e_2)+1-\e_1\e_2)^2}\right)^{\frac D2}+\ldots
\eeq
where dots denote non--leading terms. It is useful to change
parametrization of $\e_1,\e_2$ as follows
\beq
1-\e_1 =r \cos \theta\quad\quad 1-\e_2 = r \sin \theta,
\quad\quad 0\leq \theta\leq \pi/2\label{sincos}
\eeq
Then (\ref{xi1xi2'}) becomes 
\beq\label{xi1xi2''}
\left(\frac {\det(1-\Sigma\V)}{\sqrt{\det(1-S^2)}}\right)^D
\left(\frac 1{(\k+1)^2}\right)^{\frac D2}
\left(\frac 1{r^2} \frac 1{(\sin \theta+\cos\theta)^2}\right)^{\frac D2}+\ldots
\eeq
The function $(\sin \theta+\cos\theta)^{-2}$ varies between 1 and $1/2$,
with a minimum at $\theta=\pi/4$, which corresponds to $\e_1=\e_2=\e$ 
($r=\sqrt{2}(1-\e)$),
and maxima at $\theta=0,\pi/2$, which correspond to $\e_1=r, \e_2=0$
and $\e_1=0, \e_2=r$. 

These are the two possibilities considered in section 5. The first 
corresponds to $\e_1=\e_2=\e$, the second corresponds to the ordered 
limit. In between there are of course infinite many possibilities, 
giving rise to different rescalings of the number $s$.
 
\subsection{Triple limit}

We discuss here the $rhs$ of eq.(\ref{limlim}).  
We start with calculating an explicit formula for 
$\langle \hat \Xi_{\e_1} |\hat \Xi_{\e_2}*\hat \Xi_{\e_3}\rangle$,
\be
&&\langle \hat \Xi_{\e_1} |\hat \Xi_{\e_2}*\hat \Xi_{\e_3}\rangle
= \langle \hat \Xi_{\e_1} |\hat \Xi_{\e_2\star\e_3}\rangle
= \left(\frac {\hat \N_{\e_1}\hat\N_{\e_2\star\e_3}}
{\sqrt{1-\hat T_{\e_1} \hat T_{\e_2\star\e_3}}}\right)^D\0\\
&&\sim \left(\frac {\Det(1-\T\M)}{\sqrt{1-T^2}}\right)^D 
\frac 1{(1+\k)^D} \left(1+\k -(\e_1+\e_2\star\e_3)\k + \e_1 \e_2\star\e_3(\k-1)
\right)^{-D}\0\\
&&\sim \left(\frac {\Det(1-\T\M)}{\sqrt{1-T^2}}\right)^D 
\frac 1{(1+\k)^D}\0\\
&&\cdot\left(\k^2(1-\e_1)(1-\e_2)(1-\e_3)+\k(2-\e_1-\e_2-\e_3+\e_1\e_2\e_3)+
1-\e_1\e_2\e_3\right)^{-D}\label{triple1}
\ee
where we have kept only the dominant term for $\e_1,\e_2,\e_3$ near 1.
Now let us introduce the parametrization
\beq
1-\e_1= r\,\cos\theta,\quad\quad 1-\e_2= r\,\sin\theta\,\cos\varphi,
\quad\quad 1-\e_2= r\,\sin\theta\,\sin\varphi\0
\eeq
where $0\leq \theta\leq \pi/2$, $0\leq\varphi\leq \pi/2$.
Then (\ref{triple1}) becomes (keeping only the dominant term)
\be
&&\langle \hat \Xi_{\e_1} |\hat \Xi_{\e_2}*\hat \Xi_{\e_3}\rangle
\label {triple2}\\
&&\sim \left(\frac {\Det(1-\T\M)}{\sqrt{1-T^2}}\right)^D 
\frac 1{(1+\k)^D}\left(\frac 1{r^2(\cos\theta+\sin\theta(\cos\varphi
+\sin\varphi))^2}\right)^{\frac D2}\0
\ee
The function $\frac 1{(\cos\theta+\sin\theta(\cos\varphi
+\sin\varphi))^2}$ varies between a minimum of 1/3, when $\e_1=\e_2=\e_3=\e$
($r=\sqrt{3}(1-\e)$),
and a maximum of 1. Thus it is clear that for $\e_1=\e_2=\e_3$
eq.(\ref{weakeom}) cannot be satisfied. 
On the other hand there are many ways to satisfy 
\beq
\frac 1{\cos\theta+\sin\theta(\cos\varphi+\sin\varphi)}
=\frac 1{\cos \phi + \sin\phi}\0
\eeq 
in which case (\ref{triple2}) reduces to (\ref{xi1xi2''}).
The simplest way is to set $\varphi=0, \theta = \phi$ or
$\theta =\pi/2, \varphi=\phi$. These correspond to ordered limits.

\section{Derivation of ghost product}

Here we sketch a derivation of the $*_g$ product of two states of the
form (\ref{ghdsl}) given in eq.\ (\ref{ghspds}). We need to calculate
\begin{equation} \label{aghsp}
|\widetilde{\Xi}_{\tilde{\epsilon}}\rangle 
*_g |\widetilde{\Xi}_{\tilde{\eta}}\rangle
= \, _1\langle\widetilde{\Xi}_{\tilde{\epsilon}}|\, 
_2\langle\widetilde{\Xi}_{\tilde{\eta}}
|\widetilde{V}_3\rangle \;,
\end{equation}
where the ghost part of the 3-strings vertex $|\widetilde{V}_3\rangle$
is given in (\ref{gh3vert}). Using the rules for $bpz$-conjugation we
obtain
\begin{equation}
\langle \widetilde{\Xi}_{\tilde{\epsilon}} |
= \widetilde{N}_{\tilde{\epsilon}} \, \langle0|c_1^\dagger
e^{-c \widetilde{S}_{\tilde{\epsilon}} b} \;.
\end{equation}
Plugging this and (\ref{gh3vert}) in (\ref{aghsp}), and following the
steps outlined in \cite{HKw}, one obtains
\begin{eqnarray} \label{along}
\lefteqn{|\widetilde{\Xi}_{\tilde{\epsilon}}\rangle 
*_g |\widetilde{\Xi}_{\tilde{\eta}}\rangle =
\widetilde{\N}_{\tilde{\epsilon}} \, \widetilde{\N}_{\tilde{\eta}} \,\,
\det(1-\widetilde{\mathcal{T}}_{\tilde{\epsilon}\tilde{\eta}}
 \widetilde{\mathcal{M}})} \\ && \times
\left\{ 1 + c^\dagger \left[ \widetilde{\mathbf{v}}_0 
 + \left(\widetilde{V}_+,\widetilde{V}_-\right)
 (1-\widetilde{\mathcal{T}}_{\tilde{\epsilon}\tilde{\eta}}
 \widetilde{\mathcal{M}})^{-1}
 \widetilde{\mathcal{T}}_{\tilde{\epsilon}\tilde{\eta}}
 \left( \begin{array}{c} 
 \widetilde{\mathbf{v}}_+ \\ \widetilde{\mathbf{v}}_- 
 \end{array} \right) \right] b_0 \right\}
e^{c^\dagger C 
 \widetilde{T}_{\tilde{\epsilon}} * \widetilde{T}_{\tilde{\eta}}
 b^\dagger}
c_0 c_1 |0\rangle \;. \nonumber
\end{eqnarray}
Here the summations over mode indexes with positive values are
understood.
$\widetilde{\mathcal{T}}_{\tilde{\epsilon}\tilde{\eta}}$ and
$\widetilde{\mathcal{M}}$ are defined as in (\ref{hatT12})
and (\ref{MCNu}), but with tildes. We also define
\begin{equation}
\widetilde{T}_{\tilde{\epsilon}} * \widetilde{T}_{\tilde{\eta}} 
= \widetilde{X} + \left(\widetilde{X}_+,\widetilde{X}_-\right)
(1-\widetilde{\mathcal{T}}_{\tilde{\epsilon}\tilde{\eta}}
 \widetilde{\mathcal{M}})^{-1}
\widetilde{\mathcal{T}}_{\tilde{\epsilon}\tilde{\eta}}
\left( \begin{array}{c} 
 \widetilde{X}_- \\ \widetilde{X}_+
\end{array} \right) \;.
\end{equation}
Now one should observe that this formula is the same as
(\ref{hatT1T2}), and as tilded matrices satisfy the same
algebraic relations as the untilded ones , the result has 
the same form, i.e., we obtain
\begin{equation}
\widetilde{T}_{\tilde{\epsilon}} * \widetilde{T}_{\tilde{\eta}} 
= \widetilde{T}_{\tilde{\epsilon} \star \tilde{\eta}} \;.
\end{equation}
This constitutes, after using (\ref{1-T12M}), the proof of eq. 
(\ref{ghrpds}) for the reduced star product.

Now we must consider the part including the $b_0$ mode. After lengthier 
but straightforward manipulations, using the formulas from appendix A and
sections 2, 3 and 4, we obtain that the expression inside square
brackets in (\ref{along}) can be written as
\begin{equation}
\widetilde{\mathbf{v}}_0 
 + \left(\widetilde{V}_+,\widetilde{V}_-\right)
 (1-\widetilde{\mathcal{T}}_{\tilde{\epsilon}\tilde{\eta}}
 \widetilde{\mathcal{M}})^{-1}
 \widetilde{\mathcal{T}}_{\tilde{\epsilon}\tilde{\eta}}
 \left( \begin{array}{c} 
 \widetilde{\mathbf{v}}_+ \\ \widetilde{\mathbf{v}}_- 
 \end{array} \right)
= C (1-\widetilde{T}_{\tilde{\epsilon} \star \tilde{\eta}}) \mathbf{f}
\;,
\end{equation}
where $\mathbf{f}=\{f_n\}$ is the vector defined in appendix A. Using 
this in (\ref{along}) we get
\begin{equation} \label{aghpa}
| \widetilde{\Xi}_{\tilde{\epsilon}} \rangle *_g
| \widetilde{\Xi}_{\tilde{\eta}} \rangle =
\frac{\widetilde{\mathcal{N}}_{\tilde{\epsilon}} \,
\widetilde{\mathcal{N}}_{\tilde{\eta}}}{
\widetilde{\mathcal{N}}_{\tilde{\epsilon} \star \tilde{\eta}}}
\left[\frac{1+(1-\tilde{\epsilon})(1-\tilde{\eta})
\widetilde{\kappa}}{\widetilde{\kappa}+1} \right]^2
\det(1-\widetilde{T}\widetilde{\mathcal{M}}) 
\left[ c_0 + c^\dagger C
(1-\widetilde{T}_{\tilde{\epsilon} \star \tilde{\eta}}) \mathbf{f}
\right]
|\widetilde{\Xi}_{\tilde{\epsilon} \star \tilde{\eta}}\rangle \;.
\end{equation}

On the other hand, one easily shows that
\begin{equation}
\mathcal{Q}\, |\widetilde{\Xi}_{\tilde{\epsilon}}\rangle
= \left( c_0 + 
\sum_{n=1}^\infty f_n (c_n + (-1)^n c_n^\dagger) \right)
 |\widetilde{\Xi}_{\tilde{\epsilon}}\rangle \nonumber \\
= \left[c_0 + c^\dagger C (1-\widetilde{T}_{\tilde{\epsilon}})\mathbf{f}\right]
 |\widetilde{\Xi}_{\tilde{\epsilon}}\rangle \;.
\end{equation}
Using this in (\ref{aghpa}) finally one gets (\ref{ghspds}).



\begin{thebibliography}{99}

\bibitem{Ras}
L.~Rastelli, A.~Sen and B.~Zwiebach,
{\it ``Vacuum string field theory,''}
arXiv:hep-th/0106010.

\bibitem{W1} E.Witten, {\it Noncommutative Geometry and String Field
Theory},
Nucl.Phys. {\bf B268} (1986) 253.


\bibitem{Senroll}
A.~Sen,
{\it ``Rolling tachyon,''}
JHEP {\bf 0204} (2002) 048
[arXiv:hep-th/0203211].
\\
A.~Sen,
{\it``Tachyon matter,''}
JHEP {\bf 0207} (2002) 065
[arXiv:hep-th/0203265].
\\
A.~Sen,
{\it``Time evolution in open string theory,''}
JHEP {\bf 0210} (2002) 003
[arXiv:hep-th/0207105].
\\
A.~Sen,
{\it ``Time and tachyon,''}
Int.\ J.\ Mod.\ Phys.\ A {\bf 18} (2003) 4869
[arXiv:hep-th/0209122].
\\
A.~Sen,
{\it ``Open and closed strings from unstable D-branes,''}
arXiv:hep-th/0305011.
\\
A.~Sen,
{\it ``Open-closed duality: Lessons from matrix model,''}
[arXiv:hep-th/0308068].
\\
D.~Gaiotto, N.~Itzhaki and L.~Rastelli,
{\it``Closed strings as imaginary D-branes,''}
[arXiv:hep-th/0304192].
\\
I.~R.~Klebanov, J.~Maldacena and N.~Seiberg,
{\it``D-brane decay in two-dimensional string theory,''}
JHEP {\bf 0307} (2003) 045
[arXiv:hep-th/0305159].
\\
N.~Lambert, H.~Liu and J.~Maldacena,
{\it ``Closed strings from decaying D-branes,''}
arXiv:hep-th/0303139.



\bibitem{Kluson}
J.~Kluson,
{\it ``Time dependent solution in open bosonic string field theory,''}
arXiv:hep-th/0208028.
M.~Fujita and H.~Hata,
{\it ``Time dependent solution in cubic string field theory,''}
JHEP {\bf 0305} (2003) 043
[arXiv:hep-th/0304163].


\bibitem{KP} V.A.Kostelecky and R.Potting, {\it Analytical construction
of a nonperturbative vacuum for the open bosonic string},
Phys.\ Rev.\ D {\bf 63} (2001) 046007
[hep-th/{0008252}].

\bibitem{RSZ1} L.Rastelli, A.Sen and B.Zwiebach, {\it String field
theory around the tachyon vacuum},
Adv.\ Theor.\ Math.\ Phys.\  {\bf 5} (2002) 353
[hep-th/{0012251}].

\bibitem{RSZ2} L.Rastelli, A.Sen and B.Zwiebach, {\it Classical
solutions in string field theory around the tachyon vacuum},
Adv.\ Theor.\ Math.\ Phys.\  {\bf 5} (2002) 393 [hep-th/{0102112}].

\bibitem{RSZ3} L.Rastelli, A.Sen and B.Zwiebach, {\it Half-strings,
Projectors, and Multiple D-branes in Vacuum String Field Theory},
JHEP {\bf 0111} (2001) 035 [hep-th/{0105058}].

\bibitem{okuda} T.Okuda, {\it The Equality of Solutions in Vacuum String Field
Theory}, Nucl.\ Phys.\  {\bf B641} (2002) 393 [hep-th/{0201149}].

\bibitem{GRSZ2} D.~Gaiotto, L.~Rastelli, A.~Sen and B.~Zwiebach,
{\it Star Algebra Projectors}, JHEP {\bf 0204} (2002) 060 [hep-th/{0202151}].
\bibitem{MT} G.Moore and W.Taylor {\it The singular geometry of
the sliver}, JHEP {\bf 0201} (2002) 004 [hep-th/{0111069}].

\bibitem{Moeller}
N.~Moeller,
{\it ``Some exact results on the matter star-product in the half-string 
formalism,''} JHEP {\bf 0201} (2002) 019
[arXiv:hep-th/0110204].



\bibitem{HKw} H.Hata and T.Kawano, {\it Open string states around
a classical solution in vacuum string field theory},
JHEP {\bf 0111} (2001) 038 [hep-th/{0108150}].

\bibitem{HM1} H.Hata and S.Moriyama, {\it Observables as Twist
Anomaly in Vacuum String Field Theory}, JHEP {\bf 0201} (2002) 042
[hep-th/{0111034}].

\bibitem{HM2}
H.~Hata, S.~Moriyama and S.~Teraguchi,
{\it ``Exact results on twist anomaly,''}
JHEP {\bf 0202} (2002) 036
[arXiv:hep-th/0201177].

\bibitem{HM3} H.Hata and S.Moriyama, {\it Reexamining Classical Solution
and Tachyon Mode in Vacuum String Field Theory},
Nucl.\ Phys.\ B {\bf 651} (2003) 3 [hep-th/{0206208}].

\bibitem{RSZ4} L.Rastelli, A.Sen and B.Zwiebach, {\it A note on a
Proposal for the Tachyon State in Vacuum String Field Theory},
JHEP {\bf 0202} (2002) 034 [hep-th/{0111153}].

\bibitem{HK} H.Hata and H.Kogetsu {\it Higher Level Open String States
from Vacuum String Field Theory},
JHEP {\bf 0209} (2002) 027,
 [hep-th/{0208067}].

\bibitem{David} J.R.David, {\it Excitations on wedge states and on
the sliver}, JHEP {\bf 0107} (2001) 024 [hep-th/{0105184}].

\bibitem{Oka} Y.Okawa, {\it Open string states and D--brane tension
form vacuum string field theory}, JHEP {\bf 0207} (2002) 003
[hep-th/{0204012}].

\bibitem{GRSZ1} D.Gaiotto, L.Rastelli, A.Sen and B.Zwiebach,
{\it Ghost Structure and Closed Strings in Vacuum String Field
Theory}, [hep-th/{0111129}].

\bibitem{RSZ5} L.Rastelli, A.Sen and B.Zwiebach, {\it Star Algebra
Spectroscopy}, JHEP {\bf 0203} (2002) 029 [hep-th/{0111281}].


\bibitem{Sch1} M.Schnabl, {\it Wedge states in string field theory},
JHEP {\bf 0301} (2003) 004,
[hep-th/{0201095}].
{\it Anomalous reparametrizations and butterfly states in string
field theory},
Nucl.\ Phys.\ B {\bf 649} (2003) 101,
[hep-th/{0202139}].

\bibitem{GJ1} D.J.Gross and A.Jevicki, {\it Operator Formulation
of Interacting String Field Theory}, Nucl.Phys. {\bf B283} (1987) 1.

\bibitem{GJ2} D.J.Gross and A.Jevicki, {\it Operator Formulation
of Interacting String Field Theory, 2}, Nucl.Phys. {\bf B287} (1987) 225.

\bibitem{Ohta}
N.~Ohta,
{\it ``Covariant Interacting String Field Theory In The Fock Space Representation,''}
Phys.\ Rev.\ D {\bf 34} (1986) 3785
[Erratum-ibid.\ D {\bf 35} (1987) 2627].

\bibitem{tope}
L.~Bonora, C.~Maccaferri, D.~Mamone and M.~Salizzoni,
{\it ``Topics in string field theory,''}
arXiv:hep-th/0304270.

\bibitem{leclair1} A.Leclair, M.E.Peskin, C.R.Preitschopf, 
{\it String Field
Theory on the Conformal Plane. (I) Kinematical Principles},
Nucl.Phys. {\bf B317} (1989) 411.

\bibitem{leclair2} A.Leclair, M.E.Peskin, C.R.Preitschopf, {\it String 
Field Theory on the Conformal Plane. (II) Generalized Gluing},
Nucl.Phys. {\bf B317} (1989) 464.

\bibitem{Samu}
S.~Samuel,
{\it The Ghost Vertex In E. Witten's String Field Theory},
Phys.\ Lett.\ B {\bf 181} (1986) 255.

\bibitem{Oku3} K.Okuyama, {\it Ratio of Tensions from Vacuum String
Field Theory},
JHEP {\bf 0203} (2002) 050
[hep-th/{0201136}].

\bibitem{Oku1} K.Okuyama, {\it Siegel Gauge in Vacuum String Field
Theory}, JHEP {\bf 0201} (2002) 043 [hep-th/{0111087}].
I.Kishimoto, {\it Some properties of string field
algebra}, JHEP {\bf 0112} (2001) 007 [hep-th/{0110124}].

\bibitem{MM}
C.~Maccaferri and D.~Mamone,
{\it ``Star democracy in open string field theory,''}
JHEP {\bf 0309} (2003) 049
[arXiv:hep-th/0306252].


\bibitem{Oku2} K.Okuyama, {\it Ghost Kinetic Operator of Vacuum
String Field Theory}, JHEP {\bf 0201} (2002) 027 [hep-th/{0201015}].

\bibitem{belov}  D.M. Belov, A. Konechny, {\it On spectral density of Neumann 
matrices}, Phys.Lett. {\bf B558} (2003) 111-118 [hep-th/{0210169}].
D.M.Belov, {\it Diagonal Representation of Open String Star and Moyal 
Product}, [hep-th/{0204164}].
D.M.Belov, {\it Witten's Ghost Vertex Made Simple 
(bc and bosonized ghosts)}, [hep-th/{0308147}] 
E.~Fuchs, M.~Kroyter and A.~Marcus,
{\it``Virasoro operators in the continuous basis of string field theory,''}
JHEP {\bf 0211} (2002) 046
[arXiv:hep-th/0210155].

\bibitem{BMS1} L.~Bonora, D.~Mamone and M.~Salizzoni, {\it B field and
squeezed states in Vacuum String Field Theory},
Nucl.\ Phys.\  {\bf B630} (2002) 163 [hep-th/{0201060}].

\bibitem{BMS2} L.~Bonora, D.~Mamone and M.~Salizzoni, {\it Vacuum
String Field Theory with B field},
JHEP {\bf 0204} (2002) 020 [hep-th/{0203188}].

\bibitem{BMS3} L.~Bonora, D.~Mamone and M.~Salizzoni, {\it Vacuum String 
Field Theory ancestors of the GMS solitons},
JHEP {\bf 0301} (2003) 013 [hep-th/{0207044}].

\bibitem{Feng}
B.~Feng, Y.~H.~He and N.~Moeller,
{\it ``The spectrum of the Neumann matrix with zero modes,''}
JHEP {\bf 0204} (2002) 038
[arXiv:hep-th/0202176].

\bibitem{Oka2}
Y.~Okawa,
{\it ``Some exact computations on the twisted butterfly state in string 
field theory,'}'
[arXiv:hep-th/0310264.]

\bibitem{BMP2}
L.~Bonora, C.~Maccaferri, P.~Prester, in preparation


\end{thebibliography}
\end{document}